%
%

\input harvmac
\input epsf.tex

\def\figin{\epsfcheck\figin}\def\figins{\epsfcheck\figins}
\def\epsfcheck{\ifx\epsfbox\UnDeFiNeD
\message{(NO epsf.tex, FIGURES WILL BE IGNORED)}
\gdef\figin##1{\vskip2in}\gdef\figins##1{\hskip.5in}
\else\message{(FIGURES WILL BE INCLUDED)}%
\gdef\figin##1{##1}\gdef\figins##1{##1}\fi}
\def\DefWarn#1{}
\def\figinsert{\goodbreak\midinsert}
\def\ifig#1#2#3{\DefWarn#1\xdef#1{figure~\the\figno}
\writedef{#1\leftbracket figure\noexpand~\the\figno}%
\figinsert\figin{\centerline{#3}}\medskip\centerline{\vbox{
\baselineskip12pt\advance\hsize by -1truein
\noindent\footnotefont{\bf Figure~\the\figno:} #2}}
\endinsert\global\advance\figno by1}
%

\def\frac#1#2{{{#1}\over {#2}}}
\def\vev#1{{\langle {#1} \rangle}}
\def\IR{\relax{\rm I\kern-.18em R}}
\def\ie{{\it i.e.}}

\lref\SakaiCN{
  T.~Sakai and S.~Sugimoto,
  ``Low energy hadron physics in holographic QCD,''
  Prog.\ Theor.\ Phys.\  {\bf 113}, 843 (2005)
  [arXiv:hep-th/0412141].
}

\lref\AharonyUU{
  O.~Aharony, K.~Peeters, J.~Sonnenschein and M.~Zamaklar,
  ``Rho meson condensation at finite isospin chemical potential in a
  holographic model for QCD,''
  JHEP {\bf 0802}, 071 (2008)
  [arXiv:0709.3948 [hep-th]].
}

\lref\DrukkerZQ{
  N.~Drukker, D.~J.~Gross and H.~Ooguri,
  ``Wilson loops and minimal surfaces,''
  Phys.\ Rev.\  D {\bf 60}, 125006 (1999)
  [arXiv:hep-th/9904191].
}

\lref\MaldacenaIM{
  J.~M.~Maldacena,
  ``Wilson loops in large N field theories,''
  Phys.\ Rev.\ Lett.\  {\bf 80}, 4859 (1998)
  [arXiv:hep-th/9803002].
}

\lref\PestunRZ{
  V.~Pestun,
  ``Localization of gauge theory on a four-sphere and supersymmetric Wilson
  loops,''
  arXiv:0712.2824 [hep-th].
}

\lref\BrandhuberER{
  A.~Brandhuber, N.~Itzhaki, J.~Sonnenschein and S.~Yankielowicz,
  ``Wilson loops, confinement, and phase transitions in large N gauge  theories
  from supergravity,''
  JHEP {\bf 9806}, 001 (1998)
  [arXiv:hep-th/9803263].
}

\lref\AntonyanVW{
  E.~Antonyan, J.~A.~Harvey, S.~Jensen and D.~Kutasov,
  ``NJL and QCD from string theory,''
  arXiv:hep-th/0604017.
}

\lref\HeadrickCA{
  M.~Headrick,
  ``Hedgehog black holes and the Polyakov loop at strong coupling,''
  arXiv:0712.4155 [hep-th].
}

\lref\KarchSH{
  A.~Karch and E.~Katz,
  ``Adding flavor to AdS/CFT,''
  JHEP {\bf 0206}, 043 (2002)
  [arXiv:hep-th/0205236].
}

\lref\ReyIK{
  S.~J.~Rey and J.~T.~Yee,
  ``Macroscopic strings as heavy quarks in large N gauge theory and  anti-de
  Sitter supergravity,''
  Eur.\ Phys.\ J.\  C {\bf 22}, 379 (2001)
  [arXiv:hep-th/9803001].
}

\lref\AharonyDA{
  O.~Aharony, J.~Sonnenschein and S.~Yankielowicz,
  ``A holographic model of deconfinement and chiral symmetry restoration,''
  Annals Phys.\  {\bf 322}, 1420 (2007)
  [arXiv:hep-th/0604161].
}

\lref\GiveonSR{
  A.~Giveon and D.~Kutasov,
  ``Brane dynamics and gauge theory,''
  Rev.\ Mod.\ Phys.\  {\bf 71}, 983 (1999)
  [arXiv:hep-th/9802067].
}

\lref\DrukkerZQ{
  N.~Drukker, D.~J.~Gross and H.~Ooguri,
  ``Wilson loops and minimal surfaces,''
  Phys.\ Rev.\  D {\bf 60}, 125006 (1999)
  [arXiv:hep-th/9904191].
}

\lref\DeWolfePQ{
  O.~DeWolfe, D.~Z.~Freedman and H.~Ooguri,
  ``Holography and defect conformal field theories,''
  Phys.\ Rev.\  D {\bf 66}, 025009 (2002)
  [arXiv:hep-th/0111135].
}

\lref\KarchGX{
  A.~Karch and L.~Randall,
  ``Open and closed string interpretation of SUSY CFT's on branes with
  boundaries,''
  JHEP {\bf 0106}, 063 (2001)
  [arXiv:hep-th/0105132].
}

\lref\DrukkerRR{
  N.~Drukker and D.~J.~Gross,
  ``An exact prediction of N = 4 SUSYM theory for string theory,''
  J.\ Math.\ Phys.\  {\bf 42}, 2896 (2001)
  [arXiv:hep-th/0010274].
}

\lref\EricksonAF{
  J.~K.~Erickson, G.~W.~Semenoff and K.~Zarembo,
  ``Wilson loops in N = 4 supersymmetric Yang-Mills theory,''
  Nucl.\ Phys.\  B {\bf 582}, 155 (2000)
  [arXiv:hep-th/0003055].
}

\lref\DrukkerKX{
  N.~Drukker and B.~Fiol,
  ``All-genus calculation of Wilson loops using D-branes,''
  JHEP {\bf 0502}, 010 (2005)
  [arXiv:hep-th/0501109].
}

\lref\tHooftJZ{
  G.~'t Hooft,
  ``A planar diagram theory for strong interactions,''
  Nucl.\ Phys.\  B {\bf 72}, 461 (1974).
}

\lref\AharonyXZ{
  O.~Aharony, A.~Fayyazuddin and J.~M.~Maldacena,
  ``The large N limit of N = 2,1 field theories from three-branes in
  F-theory,''
  JHEP {\bf 9807}, 013 (1998)
  [arXiv:hep-th/9806159].
}

\lref\YamaguchiTQ{
  S.~Yamaguchi,
  ``Wilson loops of anti-symmetric representation and D5-branes,''
  JHEP {\bf 0605}, 037 (2006)
  [arXiv:hep-th/0603208].
}

\lref\GomisSB{
  J.~Gomis and F.~Passerini,
  ``Holographic Wilson loops,''
  JHEP {\bf 0608}, 074 (2006)
  [arXiv:hep-th/0604007].
}

\lref\WittenVV{
  E.~Witten,
  ``Current Algebra Theorems For The $U(1)$ Goldstone Boson,''
  Nucl.\ Phys.\  B {\bf 156}, 269 (1979).
}

\lref\VenezianoEC{
  G.~Veneziano,
  ``$U(1)$ Without Instantons,''
  Nucl.\ Phys.\  B {\bf 159}, 213 (1979).
}

\lref\BergmanXN{
  O.~Bergman and G.~Lifschytz,
  ``Holographic $U(1)_A$ and string creation,''
  JHEP {\bf 0704}, 043 (2007)
  [arXiv:hep-th/0612289].
}

\lref\MaldacenaRE{
  J.~M.~Maldacena,
  ``The large N limit of superconformal field theories and supergravity,''
  Adv.\ Theor.\ Math.\ Phys.\  {\bf 2}, 231 (1998)
  [Int.\ J.\ Theor.\ Phys.\  {\bf 38}, 1113 (1999)]
  [arXiv:hep-th/9711200].
}

\lref\KruczenskiME{
  M.~Kruczenski, L.~A.~P.~Zayas, J.~Sonnenschein and D.~Vaman,
  ``Regge trajectories for mesons in the holographic dual of large-N(c)  QCD,''
  JHEP {\bf 0506}, 046 (2005)
  [arXiv:hep-th/0410035].
}

\lref\PeetersFQ{
  K.~Peeters, J.~Sonnenschein and M.~Zamaklar,
  ``Holographic decays of large-spin mesons,''
  JHEP {\bf 0602}, 009 (2006)
  [arXiv:hep-th/0511044].
}

\lref\CaroneRX{
  C.~D.~Carone, J.~Erlich and M.~Sher,
  ``Extra Gauge Invariance from an Extra Dimension,''
  arXiv:0802.3702 [hep-ph].
}

\lref\BigazziJT{
  F.~Bigazzi and A.~L.~Cotrone,
  ``New predictions on meson decays from string splitting,''
  JHEP {\bf 0611}, 066 (2006)
  [arXiv:hep-th/0606059].
}

\lref\WittenQJ{
  E.~Witten,
  ``Anti-de Sitter space and holography,''
  Adv.\ Theor.\ Math.\ Phys.\  {\bf 2}, 253 (1998)
  [arXiv:hep-th/9802150].
}

\lref\KutasovUF{
  D.~Kutasov,
  ``Introduction To Little String Theory,''
{\it Prepared for ICTP Spring School on Superstrings and Related Matters, Trieste, Italy, 2-10 Apr 2001}
}

\lref\AharonyKS{
  O.~Aharony,
  ``A brief review of 'little string theories',''
  Class.\ Quant.\ Grav.\  {\bf 17}, 929 (2000)
  [arXiv:hep-th/9911147].
}

\lref\GellMannRZ{
  M.~Gell-Mann, R.~J.~Oakes and B.~Renner,
  ``Behavior of current divergences under SU(3) x SU(3),''
  Phys.\ Rev.\  {\bf 175}, 2195 (1968).
}

\lref\GubserBC{
  S.~S.~Gubser, I.~R.~Klebanov and A.~M.~Polyakov,
  ``Gauge theory correlators from non-critical string theory,''
  Phys.\ Lett.\  B {\bf 428}, 105 (1998)
  [arXiv:hep-th/9802109].
}

\lref\ItzhakiZR{
  N.~Itzhaki, D.~Kutasov and N.~Seiberg,
  ``Non-supersymmetric deformations of non-critical superstrings,''
  JHEP {\bf 0512}, 035 (2005)
  [arXiv:hep-th/0510087].
}

\lref\LukyanovNJ{
  S.~L.~Lukyanov, E.~S.~Vitchev and A.~B.~Zamolodchikov,
  ``Integrable model of boundary interaction: The paperclip,''
  Nucl.\ Phys.\  B {\bf 683}, 423 (2004)
  [arXiv:hep-th/0312168].
}

\lref\KutasovDJ{
  D.~Kutasov,
  ``D-brane dynamics near NS5-branes,''
  arXiv:hep-th/0405058.
}

\lref\NakayamaYX{
  Y.~Nakayama, Y.~Sugawara and H.~Takayanagi,
  ``Boundary states for the rolling D-branes in NS5 background,''
  JHEP {\bf 0407}, 020 (2004)
  [arXiv:hep-th/0406173].
}

\lref\ElitzurPQ{
  S.~Elitzur, A.~Giveon, D.~Kutasov, E.~Rabinovici and G.~Sarkissian,
  ``D-branes in the background of NS fivebranes,''
  JHEP {\bf 0008}, 046 (2000)
  [arXiv:hep-th/0005052].
}

\lref\LukyanovBF{
  S.~L.~Lukyanov and A.~B.~Zamolodchikov,
  ``Dual form of the paperclip model,''
  Nucl.\ Phys.\  B {\bf 744}, 295 (2006)
  [arXiv:hep-th/0510145].
}

\lref\KutasovRR{
  D.~Kutasov,
  ``Accelerating branes and the string / black hole transition,''
  arXiv:hep-th/0509170.
}

\lref\SahakyanCQ{
  D.~A.~Sahakyan,
  ``Comments on D-brane dynamics near NS5-branes,''
  JHEP {\bf 0410}, 008 (2004)
  [arXiv:hep-th/0408070].
}

\lref\AntonyanQY{
  E.~Antonyan, J.~A.~Harvey and D.~Kutasov,
  ``The Gross-Neveu model from string theory,''
  Nucl.\ Phys.\  B {\bf 776}, 93 (2007)
  [arXiv:hep-th/0608149].
}

\lref\AntonyanPG{
  E.~Antonyan, J.~A.~Harvey and D.~Kutasov,
  ``Chiral symmetry breaking from intersecting D-branes,''
  Nucl.\ Phys.\  B {\bf 784}, 1 (2007)
  [arXiv:hep-th/0608177].
}

\lref\AldayHR{
  L.~F.~Alday and J.~M.~Maldacena,
  ``Gluon scattering amplitudes at strong coupling,''
  JHEP {\bf 0706}, 064 (2007)
  [arXiv:0705.0303 [hep-th]].
}

\lref\KomargodskiER{
  Z.~Komargodski and S.~S.~Razamat,
  ``Planar quark scattering at strong coupling and universality,''
  JHEP {\bf 0801}, 044 (2008)
  [arXiv:0707.4367 [hep-th]].
}

\lref\McGreevyKT{
  J.~McGreevy and A.~Sever,
  ``Quark scattering amplitudes at strong coupling,''
  arXiv:0710.0393 [hep-th].
}

\lref\BergmanPM{
  O.~Bergman, S.~Seki and J.~Sonnenschein,
  ``Quark mass and condensate in HQCD,''
  JHEP {\bf 0712}, 037 (2007)
  [arXiv:0708.2839 [hep-th]].
}

\lref\DharBZ{
  A.~Dhar and P.~Nag,
  ``Sakai-Sugimoto model, Tachyon Condensation and Chiral symmetry Breaking,''
  JHEP {\bf 0801}, 055 (2008)
  [arXiv:0708.3233 [hep-th]].
}

\lref\ItzhakiDD{
  N.~Itzhaki, J.~M.~Maldacena, J.~Sonnenschein and S.~Yankielowicz,
  ``Supergravity and the large N limit of theories with sixteen
  supercharges,''
  Phys.\ Rev.\  D {\bf 58}, 046004 (1998)
  [arXiv:hep-th/9802042].
}

\lref\CallanAT{
  C.~G.~Callan, J.~A.~Harvey and A.~Strominger,
  ``Supersymmetric string solitons,''
  arXiv:hep-th/9112030.
}

\lref\BerensteinIJ{
  D.~E.~Berenstein, R.~Corrado, W.~Fischler and J.~M.~Maldacena,
  ``The operator product expansion for Wilson loops and surfaces in the  large
  N limit,''
  Phys.\ Rev.\  D {\bf 59}, 105023 (1999)
  [arXiv:hep-th/9809188].
}

\lref\GrossGK{
  D.~J.~Gross and H.~Ooguri,
  ``Aspects of large N gauge theory dynamics as seen by string theory,''
  Phys.\ Rev.\  D {\bf 58}, 106002 (1998)
  [arXiv:hep-th/9805129].
}

\lref\WittenZW{
  E.~Witten,
  ``Anti-de Sitter space, thermal phase transition, and confinement in  gauge
  theories,''
  Adv.\ Theor.\ Math.\ Phys.\  {\bf 2}, 505 (1998)
  [arXiv:hep-th/9803131].
}

\lref\CaseroAE{
  R.~Casero, E.~Kiritsis and A.~Paredes,
  ``Chiral symmetry breaking as open string tachyon condensation,''
  Nucl.\ Phys.\  B {\bf 787}, 98 (2007)
  [arXiv:hep-th/0702155].
}


\Title{\vbox{\baselineskip12pt\hbox{}
\hbox{WIS/06/08-MAR-DPP}}} {\vbox{ \vskip -5cm
{\centerline{Holographic Duals of Long Open Strings}}}}

\vskip  -5mm
 \centerline{Ofer Aharony$^{a}$ and David Kutasov$^b$}

\bigskip
{\sl
\centerline{$^a$Department of Particle Physics}
\centerline{The Weizmann Institute of Science, Rehovot 76100, Israel}
\centerline{\tt Ofer.Aharony@weizmann.ac.il}
\medskip
\centerline{$^{b}$EFI and Department of Physics}
\centerline{University of Chicago, 5640 S. Ellis Ave., Chicago, IL 60637, USA}
\centerline{\tt dkutasov@uchicago.edu}
\bigskip
\bigskip \medskip
}


\leftskip 8mm  \rightskip 8mm \baselineskip14pt \noindent
%
We study the holographic map between long open strings, which stretch between
$D$-branes separated in the bulk space-time, and operators in the dual boundary
theory. We focus on a generalization of the Sakai-Sugimoto holographic model
of QCD, where the simplest chiral condensate involves an operator of this type.
Its expectation value is dominated by a semi-classical string worldsheet, as for
Wilson loops. We also discuss the deformation of the model by this operator, and
in particular its effect on the meson spectrum. This deformation can be thought
of as a generalization of a quark mass term to strong coupling. It leads to the
first top-down holographic model of QCD with a non-Abelian chiral symmetry which
is both spontaneously and explicitly broken, as in QCD. Other examples we study
include half-supersymmetric open Wilson lines, and systems of $D$-branes ending
on $NS5$-branes, which can be analyzed using worldsheet methods.
\bigskip\medskip

\leftskip 0mm  \rightskip 0mm
 \Date{\hskip 8mm March 2008}

\newsec{Introduction and summary}

Holographic dualities (generalizing the AdS/CFT correspondence \MaldacenaRE) have
proven to be very useful, both for studying quantum gravity in backgrounds with
appropriate boundaries, and for studying the dual theories living on these
boundaries. However, the dictionary between the boundary theories and the  corresponding
quantum gravity duals is not yet complete.

There are two types of objects which we know how to translate between the bulk
and boundary theories (at least in the limit in which the bulk geometry is weakly curved,
and thus well described by supergravity). Local fields in the bulk map to local operators
in the dual boundary theory; sources for these fields map to sources for the
corresponding operators  \refs{\GubserBC,\WittenQJ}. Extended $p$-dimensional branes in
the bulk can end on closed $(p-1)$-dimensional surfaces on the boundary. They
correspond to operators in the boundary theory that are associated with these surfaces.

For example, when the boundary theory is a large $N$ gauge theory, a Euclidean closed
fundamental string worldsheet in the bulk, which ends on a closed loop $C$ on the
boundary, maps to (a locally supersymmetric version of) a Wilson loop in the dual field
theory \refs{\MaldacenaIM\ReyIK-\DrukkerZQ}. The latter can be thought of as associated
with external (infinitely massive) $W$-bosons in the gauge theory.

In this note we add another entry to this dictionary. When the bulk background includes
branes extending to the boundary, it is possible for other branes to end on these
branes, and give additional observables in the theory. We will focus on the case where
the bulk contains $D$-branes, and the branes ending on them are fundamental strings, but
the discussion can be generalized to other systems.

String theory in the bulk contains in this case operators corresponding to open strings
stretched between $D$-branes near the boundary. There are two qualitatively
different classes of such operators. One corresponds to strings which can shrink to zero
size (``short strings''). These are very similar to the closed string operators mentioned
above; their duals in the boundary theory are local operators, which contain the degrees of
freedom associated with the $D$-branes. The second class corresponds to ``long strings,''
that are stretched between $D$-branes which are separated by a finite amount near the
boundary. Such operators depend on the choice of an open contour $\tilde C$, which ends
on the two $D$-branes on the boundary. We propose that their duals in the boundary theory
are certain ``line operators.''

In the case of large $N$ gauge theories, $D$-branes ending on the boundary are associated
with fields in the fundamental representation of $SU(N)$, and the line operators are open
Wilson lines  starting from a field in the fundamental representation associated with the
first $D$-brane, and ending on a field in the anti-fundamental representation associated
with the second one. We propose that an insertion of such an open Wilson line in the field
theory corresponds in the bulk to an insertion of an open string ending on the corresponding
contour on the boundary, as in the Wilson loop case. As there, some correlation functions
of these operators are dominated by semi-classical string worldsheets with the appropriate
boundary conditions, and can thus be computed in the supergravity limit.

A case in which long open string operators play an important role is the Sakai-Sugimoto
holographic model of QCD \SakaiCN, and its generalizations studied in \refs{\AntonyanVW,\AharonyDA}.
This model shares with QCD the phenomena of confinement and non-Abelian chiral symmetry breaking.
As we discuss below, open Wilson lines play an important role in understanding the latter. Previous
attempts to study them in this model appeared in \refs{\BergmanPM,\DharBZ}, but our methods are different.

The Sakai-Sugimoto model describes a $4+1$ dimensional $SU(N_c)$ maximally supersymmetric
Yang-Mills (SYM) theory, with 't Hooft coupling $\lambda_5$, compactified
on a circle of radius $R$ ($x^4 \equiv x^4 + 2 \pi R$) with anti-periodic boundary conditions for
the fermions, and coupled to $N_f$ left and right-handed fermions in the fundamental representation
of $SU(N_c)$ localized at $x^4=-L/2$ and $x^4=L/2$, respectively.

The three parameters with dimensions of length, $\lambda_5$, $R$, and $L$,  can be thought of
as providing an overall scale and two dimensionless couplings  on which the dynamics depends
\refs{\AntonyanVW,\AharonyDA}. In the region of parameter space $\lambda_5\ll L\sim R$, the
$4+1$ dimensional gauge theory is weakly coupled at the scale $L\sim R$, and the model is
equivalent at long distances (much larger than $L$,  $R$, which can be viewed from this
perspective as a UV cutoff), to massless $3+1$ dimensional QCD.

For large $\lambda_5$, the $4+1$ dimensional gauge theory is strongly coupled and needs to be
UV completed. In string theory this is achieved by realizing the gauge theory as a low energy
theory on a stack of $N_c$ $D4$-branes wrapped around the twisted $x^4$ circle, intersecting
$N_f$ $D8$ and $\bar D8$-branes along an $\IR^{3,1}$,  at $x^4=\pm L/2$.

At strong coupling (and large $N_c$), one can replace the $D4$-branes by their near-horizon
geometry, and study the dynamics of the eightbranes in this geometry.  One finds that in the
vacuum the $U(N_f)_L\times U(N_f)_R$ global chiral symmetry associated with the $D8$ and
$\bar D8$-branes is spontaneously broken to the diagonal $U(N_f)$, due to the fact that the
eightbranes connect in the bulk. To study this breaking in more detail, one would like to identify an
operator in the field theory that transforms non-trivially under $U(N_f)_L\times U(N_f)_R$,
and has a non-zero vacuum expectation value (VEV) that preserves the diagonal $U(N_f)$, \ie\ an
order parameter for the symmetry breaking.

Since the left and right-handed fermions are separated  in $x^4$,  there are no local
gauge-invariant operators in the $D4$-brane theory that are charged under both $U(N_f)$ groups.
The simplest operators with the desired flavor quantum numbers are open Wilson lines of the
type discussed above, such as (for a specific choice of the contour $\tilde C$)
\eqn\simpleowzero{OW_i^j(x^{\mu})=
\psi^{\dagger j}_L(x^{\mu}, x^4=-{L\over 2}) {\cal P} \exp\left[\int_{-L/2}^{L/2} \left(i A_4 + \Phi\right) dx^4\right]
\psi_{Ri}(x^{\mu}, x^4={L\over 2})~,}
where $\Phi$ is one of the scalar fields of the SYM theory, and $\cal P$ denotes
path-ordering.

In the weak coupling regime, the gauge field $A_4$ and scalar $\Phi$ are weakly coupled at the
scale $L$, and the Wilson line in \simpleowzero\  can be neglected. Thus, the operator $OW_i^j$
reduces in this case  to the local operator $\psi^{\dagger j}_L\psi_{Ri}$, which is the familiar order
parameter of chiral symmetry breaking in field theory. In the QCD regime, its VEV is
expected to be of order $\Lambda_{QCD}^3$, where the QCD scale $\Lambda_{QCD}$ also
sets the scale of masses of mesons and glueballs in the theory.

At strong coupling, the Wilson line can not be neglected, since the $4+1$ dimensional gauge
theory degrees of freedom are strongly interacting at the scale $L$. We will compute the
expectation value of $OW_i^j$ \simpleowzero\ below and find that it is exponentially large. For
example, in the original model of \SakaiCN\ (in which $L = \pi R$), it scales like
$\exp(\lambda_5 / 18 \pi R)$. We interpret this exponentially large value as associated with
the Wilson line contribution to \simpleowzero, rather than with the fermions, since such
exponentially large values do not appear in the effective action of the Nambu-Goldstone bosons
(the ``pions'') and of the other mesons. Moreover, we will see that the expectation value depends
strongly on the choice of contour $\tilde C$ connecting the two intersections.

Another interesting question in the Sakai-Sugimoto model is how to give a mass to the
quarks.\foot{So far there are no top-down holographic examples of quark masses in theories with
a non-Abelian chiral symmetry.} As explained above, local operators which couple the left and
right-handed fermions are not gauge-invariant in this model. The best we can do is to add to the
Lagrangian the non-local operator \simpleowzero.  This breaks the chiral symmetry explicitly,
and in the region in parameter space in which the model reduces to QCD, becomes equivalent to
the quark mass deformation.

On the other hand, at strong coupling  where we can use supergravity, this deformation is
highly non-local and irrelevant (\ie\ it grows in the UV). At low energies it leads to a change
in the masses of the mesons, and in particular to a non-zero mass for the Nambu-Goldstone
bosons associated with the symmetry breaking. We will study this deformation to leading
order in the deformation parameter, and comment briefly on higher order effects.

In addition to this main example, we present two other examples of  long open string operators.
One involves a system of $k$ $NS5$-branes, with $N_f$ $Dp$ and $\bar Dp$-branes a distance
$L$ apart ending on them.  For a critical value of the distance, the branes and anti-branes can
connect, and form a single curved $D$-brane, the hairpin brane of
\refs{\LukyanovNJ\KutasovDJ-\NakayamaYX}. In the process, the $U(N_f)_L\times U(N_f)_R$
symmetry acting on the  $D$-branes breaks to the diagonal subgroup, as before. The long open
string stretched between the branes and anti-branes near the boundary can again be viewed as
an order parameter for the breaking. It has a non-zero VEV that can be computed in the same
way as for the generalized Sakai-Sugimoto model, and one can again study the
deformation that breaks the symmetry explicitly.

The main advantage of this example compared to the previous one is its tractability.
The near-horizon limit of the $NS5$-branes
is described by a solvable worldsheet theory (the linear dilaton conformal field theory
(CFT)), and the hairpin boundary state gives rise to a solvable boundary CFT, due to the fact
that it preserves ${\cal N}=2$ superconformal symmetry on the worldsheet. One can write down
explicitly the open string vertex operator corresponding to the long string, and compare the results
of our semi-classical analysis to those obtained from the effective action of the stretched open
strings, and to the exact solution of the worldsheet CFT.

A second example which we present briefly is of a  long open string operator in
type IIB string theory on $AdS_5\times S^5$
with $D$-brane defects, which preserves half of the supersymmetry, and  is analogous to the circular
closed Wilson loop in the $d=4$ ${\cal N}=4$ SYM theory. It is easy to construct many other examples
of supersymmetric open string operators, and it would be interesting to study them in more detail,
generalizing the studies of supersymmetric closed string operators. It would also be interesting
to understand if there is any relation (along the lines of \refs{\AldayHR\KomargodskiER-\McGreevyKT})
between open Wilson lines of the type studied here and scattering amplitudes of quarks and gluons.

The organization of this paper is as follows. We begin in section 2 with a general discussion of
open Wilson line operators, and their holographic description. In section 3 we discuss some
holographic computations of their correlation functions in the $D4-D8-{\bar D8}$ system. In
section 4 we study the deformation of the Lagrangian of this system by the operator
\simpleowzero, which explicitly breaks the chiral symmetry, to leading order in the deformation.
In section 5 we describe the system of $D$-branes ending  on $NS5$-branes. Finally,
in section 6 we
present a simple example of a supersymmetric long open string operator, and discuss cusp-like
divergences which occur in the computation of correlation functions of generic long open
string operators (both at weak and at strong coupling).

\newsec{Holographic open Wilson lines}

As mentioned in the introduction, in this paper we will discuss certain non-local observables
in the context of the AdS/CFT correspondence and its generalizations
\refs{\MaldacenaRE\GubserBC-\WittenQJ}. A class of such observables that has been widely
studied is Wilson loops in large $N$ gauge theories with only adjoint fields. Locally supersymmetric
Wilson loops in the fundamental representation dressed with scalar fields $\Phi_i$,
\eqn\susywilson{
W[C] = \tr \left\{ {\cal P} \exp\left[\oint_Cds \left(i A_{\mu}(x^{\nu}(s)) {\dot x}^{\mu}(s) +
n^i(s)\Phi_i(x^{\nu}(s)) |\dot x|(s)\right) \right] \right\}~,}
have been found \refs{\MaldacenaIM\ReyIK-\DrukkerZQ} to correspond to strings ending
on the closed boundary contour $C$ (parameterized by $x^{\nu}(s)$ in the non-compact
space-time, and by the unit  vector $\vec n(s)$ in the compact space). Thus, an insertion of the
operator $W[C]$ on the boundary corresponds in the bulk path integral to summing over
configurations which include a string worldsheet\foot{For Wilson loops in higher dimensional
representations of the gauge group, the dominant configurations do not look like strings but
rather like other branes carrying the same charges \refs{\DrukkerKX\YamaguchiTQ-\GomisSB}.}
ending on the loop $C$ on the boundary. Wilson loops with generic (or no) couplings to scalar
fields are more subtle; in particular, their correlation functions  have perimeter divergences
(unlike \susywilson) that need to be regularized. Nevertheless, the operators \susywilson\
already give a large amount of information about the theory. For instance, they can be used as
a diagnostic for confinement.

When the boundary theory is a large $N$ gauge theory with a finite number of fields in
the fundamental representation, the corresponding bulk description involves adding $D$-branes
to the gravity background created by the adjoint fields.\foot{This follows from 't Hooft's
\tHooftJZ\ mapping of Feynman diagrams to string worldsheets, in which loops of fields
in the fundamental representation correspond to holes in the worldsheet.} The gauge
symmetry on the $D$-branes corresponds to a global flavor symmetry in the dual field theory
(which may or may not be a symmetry of the vacuum). From the (bosonic or fermionic) fundamental
and anti-fundamental fields $\psi_i(x)$, ${\bar \psi}^j(x)$, one can form local gauge invariant
operators such as ${\bar \psi}^j(x) \psi_i(x)$. Such operators typically map under
holography to local fields in the bulk, arising from short open strings stretching from the
$i$'th to the $j$'th $D$-brane \AharonyXZ.

The situation is different when the $D$-branes are localized in some of the dimensions
in which the gauge theory lives, and thus give rise to defects. Examples include the
$D4 - D8 - \bar D8$ (generalized Sakai-Sugimoto) model
\refs{\SakaiCN\AntonyanVW-\AharonyDA} mentioned in the previous section,
the closely related intersecting brane systems described in \refs{\AntonyanQY,\AntonyanPG},
and the $D3 - D5$ system that corresponds to adding $2+1$-dimensional
hypermultiplets  to ${\cal N}=4$ SYM \refs{\KarchGX,\DeWolfePQ}.

In these cases there are no local gauge-invariant operators that involve fundamental fields
from different brane intersections (separated in space-time). The best one can do is to consider
generalizations of \simpleowzero,
\eqn\openwilson{
OW^j_i[{\tilde C}] = \bar\psi^j(x_j) {\cal P} \exp\left[\int_{\tilde C} ds\left(i A_{\mu}(x^{\nu}(s)) {\dot x}^{\mu}(s) +
n^k(s)\Phi_k(x^{\nu}(s)) |\dot x|(s) \right) \right] \psi_i(x_i)~,}
where $\tilde C$ is a contour between the point $x_j$ in the intersection at which the field
$\bar\psi^j$ lives, and the point $x_i$ in the intersection at which the field $\psi_i$ lives.
This contour is topologically a line segment; thus, we will refer to operators of the
form \openwilson\ as Open Wilson Lines, or OWLs.

Since locally along the contour $\tilde C$ the operator \openwilson\ looks just like \susywilson,
when $\vec n$ is a unit vector this operator is locally supersymmetric and its correlation functions
do not exhibit divergences proportional to the length of the path $\tilde C$. The holographic dual
of \openwilson\ must involve a string worldsheet ending on the open contour $\tilde C$ on the boundary.
Thus, we propose that an insertion of the operator \openwilson\ into the path integral of the boundary
gauge theory corresponds in the bulk to summing over configurations which include an
open
string worldsheet which approaches the contour $\tilde C$ at the boundary of the bulk space-time, and
near the boundary looks like a strip whose ends lie on the $i$'th and $j$'th $D$-branes.

As we will see, in some cases the computation of correlation functions of these operators
is dominated by a saddle point corresponding to a semi-classical string worldsheet, just like for
many holographic closed
Wilson loop computations. In particular, the one-point function $\langle OW_i^j\rangle$
is given to leading order in the semi-classical expansion by $\exp(-A/2\pi\alpha')$, where $A$ is
the minimal area of the worldsheet of such a  string. If a finite area string worldsheet does not exist,
the one-point function of the OWL vanishes.

A few comments about the preceding discussion are in order:
\item{(1)} Just like for other holographic operators, in order to obtain finite correlation
functions one needs to introduce a UV cutoff, and renormalize the OWL operators described above.
In particular, the string worldsheet that enters the calculation of the one-point function must
only have finite area for finite UV cutoff.
\item{(2)} When performing the bulk path integral in the presence of the open
string worldsheet, one has to include all the couplings of the string to the background
fields, such as the NS-NS $B_{\mu\nu}$ field, and the gauge fields that live on the $D$-branes.
\item{(3)} When the $i$'th and/or $j$'th $D$-branes give rise to more than one fundamental
field in the gauge theory, the distinction between the corresponding bulk operators in the
semi-classical calculation described above arises from quantization of zero modes on the
worldsheet of the string.

\noindent
The example that motivated this investigation is the Sakai-Sugimoto model of holographic QCD.
In this model, the large $N$ gauge theory lives on $D4$-branes in type IIA string theory, and
the fundamental fields are left and right-handed fermions, $\psi_L$ and $\psi_R$, which are
localized at $3+1$ dimensional defects -- the intersections of the $D4$-branes with $N_f$ $D8$
and $\bar D8$-branes, respectively. The $D8$ and $\bar D8$-branes are separated by a distance
$L$ in the direction $x^4$ along the $D4$-branes. The model has a $U(N_f)_L\times U(N_f)_R$
global symmetry corresponding to the gauge symmetry on the $D8$ and $\bar D8$-branes.

In the strongly coupled regime $\lambda_5\gg L,R$, the vacuum of this model corresponds to
a brane configuration in which the $D8$ and $\bar D8$-branes are connected, and the chiral
$U(N_f)_L\times U(N_f)_R$ symmetry is dynamically broken to the diagonal $U(N_f)$. Most of
the work on the model involved light open $8-8$ strings, such as the translational modes of
the eightbranes and their worldvolume gauge fields. The latter correspond in the boundary theory
to the $U(N_f)_L\times U(N_f)_R$ chiral symmetry currents $\psi_L^\dagger(x)\sigma^\mu\psi_L(x)$,
$\psi_R^\dagger(x)\bar\sigma^\mu\psi_R(x)$.

These ``short string'' operators are useful for analyzing the low lying spectrum of mesons,
but in order to study chiral symmetry breaking it is better to consider operators such as
\simpleowzero, which transform as $\bf(N_f,\bar N_f)$ under the chiral symmetry. In the
next section we will use
holography to show that the expectation value of these operators is non-zero at strong coupling;
thus, they are natural order parameters for chiral symmetry breaking.

In QCD one can break the chiral symmetry explicitly by adding a mass term for the quarks. The closest
analog of this at strong coupling is to add \simpleowzero\ to the Lagrangian. We will describe some
results about this deformation in section 4 below.

OWL operators of the form \openwilson\ can in principle be also defined for theories in which
the fundamental fields are not localized at defects, but they seem to be less useful in such
cases. Consider, for example, the $D3-D7$ system, which corresponds to adding to ${\cal N}=4$
SYM a massless hypermultiplet in the fundamental representation of the gauge group. In the dual description
this corresponds \refs{\AharonyXZ,\KarchSH} to adding a $D7$-brane wrapping $AdS_5\times S^3$ to type
IIB string theory on $AdS_5\times S^5$, where the $S^3$ is a maximal three-sphere inside $S^5$.

The one-point function of an open Wilson line operator \openwilson\ involves in this case a
configuration with a string ending on the contour ${\tilde C}$ connecting the points $x_i$ and $x_j$
in $\IR^{3,1}$. The worldsheet of such a string can always reduce its area by contracting towards
the boundary. Therefore, the corresponding one-point function depends on the UV regulator, and
does not appear to be well-behaved.\foot{This is also true at weak coupling, due to divergences
associated with the screening by the fundamental representation fields.} This is reasonable from
the general perspective described in the introduction. The open string in question is really a short
$D7-D7$ string that does not shrink only because its two ends are held fixed at two different points in
$\IR^{3,1}$.  A more natural basis for describing such strings is in terms of excited perturbative
$D7-D7$ strings, rather than the OWL basis \openwilson.

The last remark is also applicable to long open strings connecting widely separated $D$-branes.
For a given pair of branes there is a preferred contour $\tilde C$ that has minimal length, and
it is natural to study the OWL operator associated with it. For the $D4-D8-\bar D8$ system this is
the operator \simpleowzero. One can consider other contours that connect different points in $\IR^{3,1}$
and/or vary non-trivially in the interior, as in \openwilson, but these are less natural.
They can be alternatively described by adding string oscillators to the operator corresponding
to the minimal contour \simpleowzero.

\newsec{Open Wilson lines in the $D4-D8-{\bar D8}$ system}

To demonstrate the general discussion of the previous section, we will consider here the following
intersecting brane system in type IIA string theory. We start with $N_c$ $D4$-branes stretched in
the $\IR^{4,1}$ labeled by $(x^0,x^1,x^2,x^3,x^4)$, and add to them $N_f$ $D8$-branes localized
at $x^4=-L/2$, as well as $N_f$ $\bar D8$-branes localized at $x^4=+L/2$.

This leads \AntonyanVW\ to a non-confining theory of massless left and right-handed
fermions, $\psi_L$, $\psi_R$, which are localized at the $4-8$ and $4-\bar 8$
intersections, respectively,  and  interact via exchange of modes living on the
$D4$-branes.\foot{The Sakai-Sugimoto model \SakaiCN\ is obtained by compactifying
$x^4$ on a circle, with twisted boundary conditions for the fermions on the
$D4$-branes \WittenZW. We will comment on the generalization of our considerations
to this case below.} The strength of the interaction is determined by the 't Hooft
coupling $\lambda_5 = (2\pi)^2 g_s N_c l_s$. When the interaction at the scale $L$
is strong $(\lambda_5\gg L)$, one can replace the $D4$-branes by their near-horizon
geometry \refs{\ItzhakiDD,\WittenZW}. The metric is given by
\eqn\dfourmet{ds^2 = \left(\frac{u}{R_{D4}}\right)^{3/2} \left[-(dx^0)^2 +
\sum_{i=1}^4 (dx^i)^2\right] + \left(\frac{R_{D4}}{u}\right)^{3/2}
\left[du^2+u^2
d\Omega_4^2\right]~,}
where $R_{D4}^3 \equiv \pi g_s N_c l_s^3$. The RR four-form and dilaton are
\eqn\dfourdil{F_{(4)} = \frac{2 \pi N_c}{{\rm Vol}(S^4)} \epsilon_4,\qquad e^{\Phi} = g_s
\left(\frac{u}{R_{D4}}\right)^{3/4}~.}
The dynamics of the fermions is determined by the shape of the eightbranes
in this background. It was found in \AntonyanVW\ that in the lowest energy
configuration the $D8$ and $\bar D8$-branes are connected by a tube and form
a single stack of $N_f$ connected eightbranes. They are extended in the $\IR^{3,1}$
labeled by $(x^0, x^1,x^2, x^3)$, wrap the four-sphere labeled by $\Omega_4$, and form
a curve $u(x^4)$ in the $(u, x^4)$ plane, which is a solution of the first order differential
equation
\eqn\foru{{u^4\over \sqrt{1 + \left({R_{D4}\over u}\right)^3 u'^2}} = u_0^4~.}
The solution of \foru\ is a $U$-shaped brane, with the distance between the two arms
approaching $L$ at large $u$. The minimal value of $u$ to which the $D8$-branes extend,
$u_0$, is determined by $L$,
\eqn\uol{L = {1\over 4} R_{D4}^{3/2} u_0^{-1/2}B({9\over 16},{1\over 2})~.}
For strong coupling, the curvature of the metric \dfourmet\ near the $D8$-branes
is small. The string coupling \dfourdil\ diverges as $u\to\infty$, but in the 't Hooft
large $N_c$ limit there is a parameterically wide region in $u$ in which it is small,
and we can restrict attention to that region by placing a UV cutoff on $u$, $u\le u_{\rm max}$.

Since the $D8$ and $\bar D8$-branes are connected in the vacuum, the chiral
$U(N_f)_L\times U(N_f)_R$ symmetry acting on them is spontaneously broken to its diagonal
subgroup. The fermions $\psi_L$, $\psi_R$, which correspond in the brane
picture to strings stretching from the bottom of the curved $D8$-branes towards
$u=0$, obtain a dynamically generated ``constituent mass'' $m=u_0/2\pi\alpha'\sim \lambda_5/L^2$.

One can use the effective action for the $D8$-branes (which includes the DBI, Wess-Zumino, and
fermionic terms) to study the low-lying excitations of the model,  and in particular to verify
the existence of $N_f^2$ massless Nambu-Goldstone mesons corresponding to the breaking of the
chiral symmetry. The rest of the spectrum is massive; the masses of the lowest lying mesons are
of order $1/L$. They are much lighter than the fermions, and can be thought of as tightly bound
states of two fermions.

As explained above, the simplest operator which can serve as an order parameter
for chiral symmetry breaking in this theory is the OWL operator \simpleowzero. We next turn
to a calculation of its one-point function at strong coupling, the chiral condensate,
and comment on more general operators of the form \openwilson.

\subsec{One-point functions of open Wilson lines}

In order to compute the expectation value of the OWL operator \simpleowzero\ at
strong coupling, we need to perform the gravitational path integral in the closed
string background \dfourmet, \dfourdil, in the presence of the curved $D8$-branes
described around \foru, and of a Euclidean fundamental string worldsheet which
near the (regularized) boundary at $u=u_{\rm max}$ stretches along a straight line
in the $x^4$ direction between the two arms of the curved $D8$-branes.

\medskip
\ifig\diskinst{The semi-classical worldsheet which gives the chiral condensate in the
$D4-D8-\bar D8$ model is drawn in green.}
{\epsfxsize3.2in\epsfbox{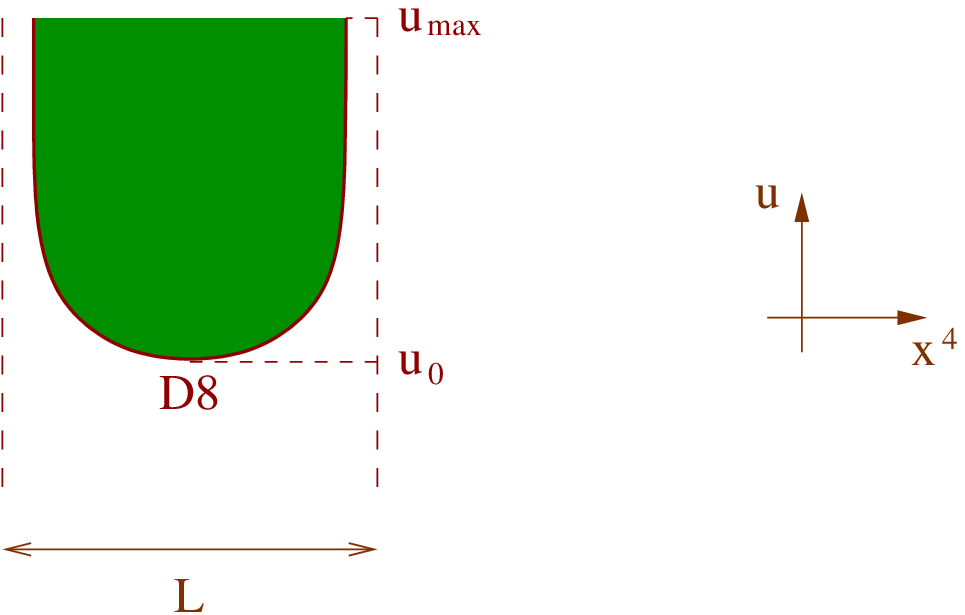}}

\noindent
This path integral is dominated by a semi-classical contribution of a Euclidean
worldsheet,  which is localized at a point in $\IR^{3,1}\times S^4$, and fills the
region in the $(u, x^4)$ plane between the boundary at $u=u_{\rm max}$ and the
$D8$-branes (see  \diskinst).

To leading order in $\alpha'$ the action of such a string is proportional to its area,
\eqn\areastraight{S_{\rm str} =
{1\over {2\pi \alpha'}} \int dx^4 \int_{u(x^4)}^{u_{\rm max}} du\sqrt{g_{uu} g_{44}}
= {1\over {2\pi \alpha'}} \int dx^4 \left[u_{\rm max} - u(x^4)\right]~.}
Performing the integral one finds
\eqn\regarea{S_{\rm str} =
{u_{\rm max}L\over2\pi\alpha'}-{R_{D4}^{3/2} u_0^{1/2} \over {8 \pi \alpha'}} B({7\over {16}},{1\over 2})
={u_{\rm max}L\over2\pi\alpha'}-C_1 {\lambda_5 \over L}~,}
where the constant $C_1$ is given by
$C_1 = B({7\over16},{1\over2}) B({9\over16},{1\over2}) / 128 \pi^2 \simeq 0.0079$,
and we neglected corrections that go to zero in the limit $u_{\rm max}\to\infty$.
The term proportional to the UV cutoff $u_{\rm max}$ on the right-hand side is independent of the
coupling $\lambda_5$, and can be absorbed in the definition of the operator
\simpleowzero.\foot{As explained in \DrukkerZQ, this term is naturally canceled by a
Legendre transform which is part of the definition of locally supersymmetric Wilson
line operators.}

Thus, we conclude that at strong coupling the expectation value of the operator \simpleowzero\
is given by
\eqn\vevwilson{
\langle OW^j_i \rangle \simeq \delta_{ij}\exp\left(-S_{\rm str}\right)\simeq\delta_{ij} \exp(C_1 \lambda_5 / L)~.}
The calculation above captures the leading behavior of this one-point function at strong coupling.
The first subleading corrections come from quadratic fluctuations around the Euclidean worldsheet
of \diskinst, and from the coupling of the string to the varying dilaton. They are expected to
give a polynomial pre-factor in front of the exponential \vevwilson.

The non-vanishing expectation value \vevwilson\ exhibits the expected pattern of chiral
symmetry breaking, $U(N_f)_L\times U(N_f)_R\to U(N_f)_{\rm diag}$, in agreement with our
earlier discussion. It grows exponentially with the coupling $\lambda_5/L$ in the region
$\lambda_5 \gg L$ in which our calculation is reliable. At first sight this might seem surprising,
since in the weakly coupled field theory regime the chiral condensate is closely related
to the dynamically generated fermion mass, whereas here this is not the case -- the fermion
mass scales like $\lambda_5/L^2$, those of the mesons scale like $1/L$, while the condensate
\vevwilson\ is exponentially large.\foot{
The chiral condensate we find is also widely separated from the pion decay constant
$f_{\pi}$, which will be discussed in the next section.}
The difference between the
two regimes is that for strong coupling most of the contribution to \vevwilson\ appears to be
due to the Wilson line in \simpleowzero\ rather than to the fermion bilinear part of the operator,
while for weak coupling this Wilson line gives a negligible contribution.

The fact that the exponential behavior of \vevwilson\ at strong coupling is due to the Wilson line
rather than to the fermions can be seen more quantitatively by studying its dependence on the
contour $\tilde C$. Consider, for instance, the one point function of an open Wilson line \openwilson\ connecting two
points in $\IR^{3,1}$, $x_0^\mu$ and $x_1^{\mu}$, (space-like) separated by a distance much larger than $L$.
A class of contours connecting these points that is useful for our purposes involves moving first in
$x^4$ (at a fixed value of $x^\mu$, $x^{\mu}=x_0^{\mu}$) from $-L/2$ to some $x_0^4$, then  varying
$x^\mu$ from $x_0^{\mu}$ to $x_1^{\mu}$ at fixed $x^4$, and finally moving again in $x^4$ to $L/2$.
Such contours have cusps, but these can be smoothed out (and in any case the divergences they lead to
are well understood and can be subtracted out).

Finding the precise shape of the string worldsheet which minimizes the action with these boundary
condition is rather complicated. However, when the two points $x_0$ and $x_1$ are widely separated,
we expect the main contribution to this expectation value to come from the part of the worldsheet
at $x^4=x^4_0$. In the special case $x_0^4=0$, this part of the  worldsheet is easy to analyze. Its
contribution to the regularized action is given by
\eqn\ssep{S_{\rm str}=-{u_0 |\vec{x}_0-\vec{x}_1| \over 2 \pi \alpha'}
\propto-{\lambda_5\over L^2} |\vec{x}_0-\vec{x}_1|~.}
The proportionality constant on the right-hand side can be read off from \uol. On the other hand,
when $x_0^4$ approaches (say) $L/2$, the regularized action turns out to be proportional to
$-\lambda_5 |\vec{x}_0-\vec{x}_1| / (x_0^4 - L/2)^2$.

We see that the expectation values of these operators, proportional to $\exp(-S_{\rm str})$, grow exponentially with the distance between
the endpoints of the contour $\tilde C$ in $\IR^{3,1}$, and the coefficient of the distance in the
exponent depends on the precise contour we choose.  We conclude that this exponential growth is a
property of the contour rather than of the fermion bilinear at its ends. This also explains why the
expectation value under consideration does not decay exponentially with the distance in $\IR^{3,1}$,
$|\vec{x}_0-\vec{x}_1|$, as one might have expected due to the fact that the fermions are massive.

So far we have discussed the computation of the chiral condensate for the extremal $D4$-brane
background, but it is easy to generalize the discussion to the case where $x^4$ lives on a
circle of radius $R$, with anti-periodic boundary conditions for the fermions. The near-horizon
$D4$-brane geometry is in this case a Wick-rotated Euclidean black hole geometry, in which the
radius of the $x^4$ circle varies between its asymptotic value $R$ at large $u$ and zero at
$u=u_{\Lambda}=\lambda_5 \alpha' /9\pi R^2$ \WittenZW.

One can again analyze the shape of the $D8$-branes as a function of $L$ and $R$ and calculate the
expectation value of OWL operators such as \simpleowzero,  by evaluating the area of the
corresponding Euclidean string worldsheet. There are some small differences in the precise form of
the solutions for the $D8$-branes and for the strings, but the qualitative properties are not modified.
There are now two independent operators of the form \simpleowzero, one with the Wilson line going in the positive $x^4$
direction and the other in the negative $x^4$ direction. For generic $L$, $R$, the worldsheets that
determine the expectation values of the two operators have different areas, so one of the operators
has a larger VEV. In the special anti-podal case $L=\pi R$ considered in \SakaiCN\ the one-point
functions of both operators are given by
\eqn\openss{
\langle OW^j_i \rangle \simeq\delta_{ij} \exp(\lambda_5 / 18 \pi R)~.}
For general $L\ll \lambda_5$ one finds a result that smoothly interpolates between \openss\ for
$L=\pi R$ and \vevwilson\ for $L \ll R$, where the modification of the background at the location
of the $D8$-branes due to the compactness of $x^4$ is negligible.

The discussion above was restricted to the strong coupling regime $\lambda_5\gg L$. In the opposite
limit, $\lambda_5 \ll L\sim R$, at energies much smaller than $1/R$ one expects the model to reduce
to QCD with massless quarks. In this limit the gauge field $A_4$ and the scalar fields are expected
to decouple \WittenZW, so \simpleowzero\ should go over to the usual chiral condensate of QCD, which
scales as $\Lambda_{QCD}^3 \simeq {1\over R^3}\exp(- 16 \pi^3 R / \lambda_5) $. If there is no phase
transition as one varies $\lambda_5/R$, we expect a smooth interpolation between this result and \openss.

For $L\ll R$, and in particular in the limit $R\to\infty$ with fixed $L$, the situation
is not completely clear. At strong coupling ($\lambda_5 \gg L$), one finds in this limit a theory which breaks chiral
symmetry but does not confine, which can be thought of as a particular UV completion of
the Nambu-Jona-Lasinio model \AntonyanVW. Field theoretic intuition suggests that at weak coupling ($\lambda_5 \ll L$)
chiral symmetry is not broken, and thus the theory undergoes a phase transition at some critical
value of the coupling $\lambda_5/L$, but this has not been conclusively established yet.

In other closely related brane systems, discussed in \refs{\AntonyanQY, \AntonyanPG}, which give rise
to $1+1$ dimensional intersections, such as the $D4-D6-{\bar D6}$ system, one can analyze the dynamics
for both weak and strong coupling, and in particular calculate the expectation value of \simpleowzero\
in both limits (for any value of $R$). The strong coupling computation is very similar to that described above,
and gives $\vev{OW} \sim \exp({\tilde C}_1 \lambda_5 / L)$ with some calculable constant ${\tilde C}_1$ that
depends on $L/R$ and approaches a finite value as $L/R\to 0$.

For weak coupling and infinite $R$, one obtains in this case a generalized Gross-Neveu model which can be analyzed
using field theoretic methods and gives $\vev{OW} \sim \exp(- L / \lambda_5)$. For finite $R$ one gets a
generalization of the 't Hooft model of two dimensional QCD that includes four-Fermi interactions, and is
solvable at large $N_c$, like its two extreme limits~-- the 't Hooft and Gross-Neveu models. It would be
interesting to compute the chiral condensate in this model as a function of $L/R$, and compare it to the strong
coupling calculation described above. We expect a smooth interpolation between the strong and weak coupling
limits as one varies the parameters $\lambda_5/L$, $\lambda_5/R$ that govern chiral symmetry breaking
and confinement, respectively.

\subsec{Correlation functions of open Wilson lines}

The computation of the expectation value of a product of several OWLs \openwilson\ is also
straightforward in principle, but in practice it is more difficult to find the
appropriate semi-classical string worldsheets (if they exist). As in correlation functions
of closed Wilson loops, in some cases a correlation function of a product of OWLs is
dominated by a single semi-classical worldsheet; in other cases it is
dominated by several semi-classical worldsheets connected by propagators in the bulk
(at leading order in $1/N_c$ they must be connected by propagators of open string fields);
in yet other cases
there may be no semi-classical contribution at all. In the supergravity limit, there can be
sharp phase transitions between the first two possibilities, as in closed Wilson loop correlators \GrossGK.

\ifig\twoow{The two semi-classical configurations that dominate the computation of $F_2$. On the
left we have the two-string configuration, with one string (as in \diskinst) ending on a $D8$-brane at $x_0$ and
the other at $x_1$, connected by an open string propagator inside
the $D8$-brane. On the right we have the single string configuration. The string worldsheets
are filled with diagonal lines, and the $D8$-brane lives everywhere but was only drawn at $x_0$ and $x_1$.}
{\epsfxsize5in\epsfbox{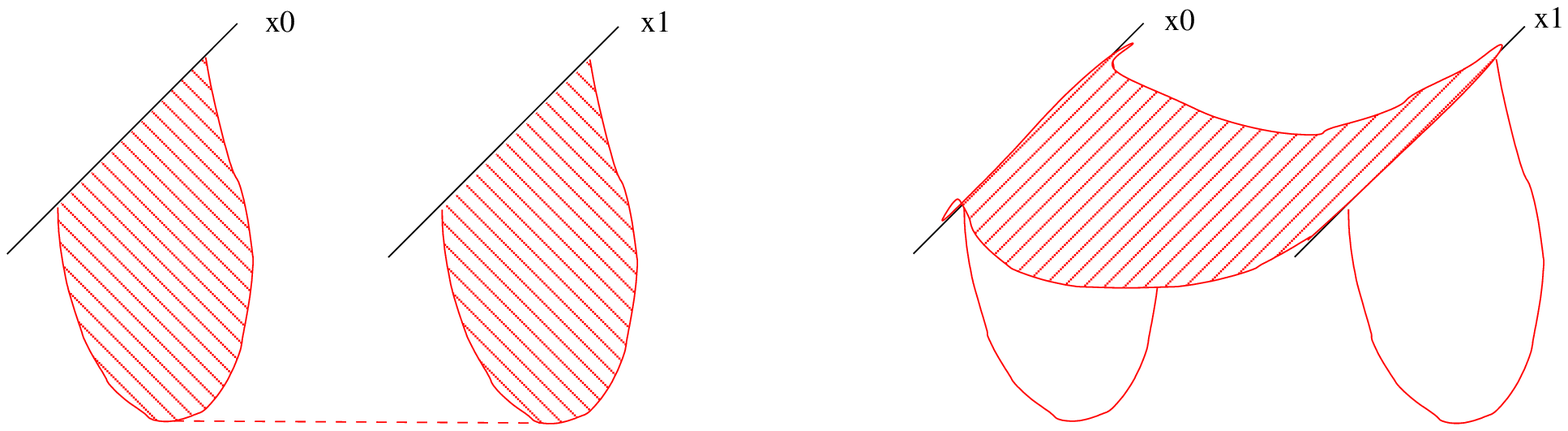}}

A case where the dominant worldsheets are easy to describe is the correlation
function $F_2 \equiv \vev{OW_i^i(x_0^{\mu}) (OW_i^i)^{\dagger}(x_1^{\mu})}$ (no sum over
$i$ implied) in the $D4-D8-{\bar D8}$ system. There are two distinct semi-classical contributions
to this correlation function. One involves the worldsheets that appear in the computation of the
one-point functions of  $OW^i_i$ and $({OW_i^i})^\dagger$ (the worldsheet corresponding to
$OW^{\dagger}$ is the same as the one for $OW$, but with an opposite orientation), connected by a
propagator of an open string field on the $D8$-brane. It is depicted in the left part of \twoow.
The leading order contribution at large distances is due to massless pion exchange, and should
be proportional to
\eqn\firstftwo{F_2 \simeq {\vev{OW_{ii}} \vev{OW_{ii}}^{\dagger} \over |\vec{x}_0 - \vec{x}_1|^2} \simeq
{\exp(2 C_1 \lambda_5 / L) \over |\vec{x}_0 - \vec{x}_1|^2}~.}

\noindent
The second semi-classical worldsheet smoothly connects the two OWL's at $x_0^{\mu}$ and $x_1^{\mu}$,
by extending into the bulk, as in the right part of \twoow.

If the $D8$ and $\bar D8$-branes were localized at fixed values of $x^4$, this configuration would be
precisely the one that appears
in the computation of the energy of a quark and an anti-quark separated by a distance
$|\vec{x}_0-\vec{x}_1|$, with $x^4$ playing the role of time (the worldsheet would simply
stretch in this direction and end on the $D8$-branes at $x^4 = \pm L/2$). In the $D4$-brane background
\dfourmet, \dfourdil, this energy is given by \BrandhuberER\ $(-\lambda_5 /|\vec{x}_0-\vec{x}_1|^2)$,
so in this case we would obtain
\eqn\secondftwo{F_2 \simeq \exp(\lambda_5 L / |\vec{x}_0-\vec{x}_1|^2)~.}
In the actual configuration we are interested in, the $D8$-branes bend in $x_4$, and \secondftwo\
should be modified by taking their shape into account. When the extent of the string
in the radial direction becomes comparable to $u_0$, this modification is significant.  However, for
short distances, \secondftwo\ is still reliable.\foot{Note that $F_2$  diverges as
$\vec{x}_0 \to \vec{x}_1$ (when the cutoff is sent to infinity).}

As the distance increases,  the area of the worldsheet in the right part of \twoow\ increases, and at some
point it becomes larger than that in the left part. At that point, the correlation function in question
makes a phase transition from \secondftwo\ to \firstftwo. This transition is expected to occur at
$|\vec{x}_0 - \vec{x}_1| \simeq L$.

An example of a correlator for which there is no obvious smooth worldsheet configuration at short
distances is $\vev{OW_{i}^i(x_0^{\mu}) OW_{i}^i(x_1^{\mu})}$. The string ending at $x_0^{\mu}$
would have to change its orientation in the bulk before coming back to end at $x_1^{\mu}$. Thus,
in this case it seems likely that the two-string configuration on the left of \twoow\ always dominates
and gives the behavior \firstftwo.

Another interesting correlator is $\vev{\det(OW_i^j(x^{\mu}))}$, which is a singlet of the non-Abelian
$SU(N_f)_L\times SU(N_f)_R$ but carries axial $U(1)$ charge. In the case of finite $R$, due to the
axial anomaly, this should be non-zero even in phases where the chiral symmetry is not spontaneously
broken and the $D8$-branes and $\bar D8$-branes do not connect. The computation of
$\vev{\det(OW_i^j)}$ involves in this case $N_f$ strings ending on the boundary, on the $D8$-branes
and on the $\bar D8$-branes. Naively it vanishes when the $D8$-branes and $\bar D8$-branes do not
connect, since the strings have nowhere to end in the IR. However, there are contributions from Euclidean $D0$-branes wrapped around the $x^4$ circle
(which are instantons from the point of view of the $4+1$ dimensional gauge theory). Such Euclidean
$D0$-branes should have $N_f$ fundamental strings ending on them between the $D8$-branes and the
$\bar D8$-branes\foot{Recall that the $D8$-branes generate a ten-form flux, which couples to the gauge
field on the $D0$-branes.} \BergmanXN. These strings can extend to the boundary and thus contribute to
$\vev{\det(OW_i^j)}$. Being $D$-instanton effects, such contributions are exponentially suppressed in the
't Hooft large $N_c$ limit, but they are the leading contribution to $\vev{\det(OW_i^j)}$ in phases where
the non-Abelian chiral symmetry is unbroken.

\newsec{Deforming by open Wilson lines}

In the previous section we computed the expectation value of the OWL operator \simpleowzero\ in the
generalized Sakai-Sugimoto model. In this section we will study a deformation of the model that
corresponds to adding this operator to the Lagrangian,
\eqn\deform{\delta S =\kappa \int d^4x \sum_{j=1}^{N_f} OW_j^j(x) + c.c.~.}
This leads to explicit breaking of the $U(N_f)_L\times U(N_f)_R$ chiral symmetry to the diagonal $U(N_f)$,
in addition to the spontaneous breaking present at $\kappa=0$. The deformation \deform\ can be thought of
as a generalization to strong coupling of a ``current mass'' for the fermions,
$\delta S=\kappa\int d^4x  \sum_j \psi^{\dagger j}_L \psi_{Rj} + c.c.$, that plays a role in QCD. The generalization to non-equal ``masses'' $\kappa_j$ for different quark flavors is
straightforward.

We will study the deformed theory semi-classically at strong coupling, in the hope that the strong
coupling results are smoothly related to large $N_c$ QCD with massive quarks. We will work to first
order in the mass parameter $\kappa$; this involves a single insertion of the perturbation, for which
we can use our results from the previous section. In QCD this is a good approximation for the $u$ and
$d$ quarks, whose current mass is much smaller than the QCD scale. It would be interesting to go beyond
first order in $\kappa$. For this, one needs to evaluate $n>1$ point functions of the operators
$OW_{i}^i(x)$, which are complicated, as discussed in the previous section.

To first order in $\kappa$, the deformation \deform\ can be described by adding to the
space-time action the term
\eqn\lamdeform{\delta S = {\kappa \over {\rm Vol}(S^4)} \int d^4x \int d^4\Omega
\sum_i e^{-S^{(i)}_{\rm str}} + c.c.~,}
where $S_{\rm str}^{(i)}$ is the action of the string ending on the $i$'th $D8$-brane discussed in the
previous section,\foot{Of course, this only makes sense in the phase in which the $D8$ and $\bar D8$-branes
are connected. In phases like the high-temperature phase of the Sakai-Sugimoto model in which the branes are
not connected, $\vev{OW}$ vanishes, and we do not have a semi-classical description of the deformation.} and
the integral over the four-sphere implements an average over the scalar field that enters the
definition of the operator \simpleowzero, restoring the $SO(5)$ symmetry of the model.

The deformation \lamdeform\ is non-local\foot{Similar non-local mass terms were also recently considered in \CaroneRX.}, since the action $S_{\rm str}^{(i)}$ depends on the position
of the $D8$-branes everywhere in the radial coordinate (it also includes a coupling to the gauge field
on the $D8$-branes, and to closed string fields). This is not surprising, since the field theory
deformation \deform\ is non-local. In the dual string description, in addition to the explicit
non-locality in the direction of the Wilson line, we also have non-locality in the radial
direction.\foot{This non-locality could be avoided if instead
of deforming by $(OW + c.c.)$ we would deform by $(\ln(OW) + c.c.)$, since this would just shift
the action by a multiple of the action $S_{\rm str}$ of the stretched string, which is an integral of a
local function of the $D8$-brane fields (a similar perturbation for closed Wilson loops was recently
considered in \HeadrickCA). However, such a deformation does not have the same symmetry
properties as \deform\ (in particular it does not break the axial $U(1)$ symmetry), and it is not
obvious that $\vev{\ln(OW)}=\ln(\vev{OW})$ semi-classically, so we will not consider it further here.}

The deformation \deform, \lamdeform, is of order $N_c$ (or $1/g_s$), like any other open string deformation,
so it is expected to influence open string fields (like the position of the $D8$-branes) at leading order,
while the corrections to closed string fields (like the metric) are suppressed by a power of $g_s$.
One thing that is relatively easy to compute is the mass of the Nambu-Goldstone bosons (the ``pions'')
due to the deformation \deform\ at leading order in $\kappa$.

For $\kappa=0$ we have a $U(N_f)_L \times U(N_f)_R$ global symmetry spontaneously broken to $U(N_f)$.
In the effective field theory on the $D8$-branes the order parameter for this breaking can be taken to
be the holonomy matrix
\eqn\holmat{U \equiv {\cal P} \exp(i \int_{-L/2}^{L/2} dx^4 {\tilde A}_{x^4})~,}
where $\tilde A$ is the gauge field on the $D8$-branes.  The matrix $U$ transforms as a bifundamental of
$U(N_f)_L\times U(N_f)_R$.\foot{Parameterizing the position of the $D8$-branes in the $(u,x^4)$ plane by
a variable $z$ which goes from minus infinity at one boundary of the branes to plus infinity at the other
boundary, we can write $U = {\cal P} \exp(i \int_{-\infty}^{\infty} dz {\tilde A}_z)$.} It is precisely the matrix
appearing in the low-energy chiral Lagrangian,\foot{Naively one might think that the holonomy matrix $U$
could serve as an order parameter for the chiral symmetry breaking in the full string theory as well.
However, while the holonomy is gauge-invariant
in the $D8$-brane gauge theory, it is not gauge-invariant in the full string theory under
gauge transformations of the NS-NS $B$ field. In order to obtain a gauge-invariant
object we must multiply $U$ by $\exp(i\int B)$ where the integral is over a surface
bounded by the $D8$-branes. The only way to construct an operator containing this phase
in string theory is to put in a fundamental string (or another object with the same
charges) ending on the $D8$-branes, giving precisely the OWL operators
discussed above. Thus, one can think of $OW$ as a completion of $U$ to the full string
theory; in the $N_f=1$ case $U$ is the phase of $OW$.}
which is usually written in terms of pion
fields as $U(x) = \exp(i \pi(x) /
f_{\pi})$. Its low-energy effective Lagrangian is given by
\eqn\lleef{L_{eff} = (f_{\pi}^2/4) \Tr(\del_{\mu} U\del^{\mu} U^{\dagger})~,}
with  \refs{\SakaiCN,\AharonyUU}
\eqn\forfpi{f_{\pi}^2 \simeq \lambda_5 N_c / L^3~.}
The $N_f^2$ pion fields in $U$ are massless Nambu-Goldstone bosons.\foot{The axial $U(1)$ symmetry is anomalous,
and the corresponding pion obtains a mass at order $1/N_c$ \refs{\WittenVV,\VenezianoEC,\SakaiCN,\BergmanXN}.}

The deformation \deform\ explicitly breaks the chiral symmetry to the diagonal subgroup, and is expected
to give a mass to all the Nambu-Goldstone bosons. Indeed, the perturbation $\exp(-S_{\rm str}^{(i)})$ in
\lamdeform\ includes a coupling to the gauge field on the $D8$-branes, of the form
$({\cal P} \exp[-i \int_{-L/2}^{L/2} dx^4 {\tilde A}_{x^4}])_{ii}$. This coupling did not play
a role in our evaluation of $\vev{OW}$, since we assumed that we were expanding
around a configuration in which the gauge field on the $D8$-branes vanishes, but it is important
in analyzing the perturbed theory \lamdeform.  The effective Lagrangian for $U$, \lleef, is deformed
at first order in $\kappa$ by
\eqn\piondeform{\delta L_{eff} = |\vev{OW}| \kappa \tr(U) + c.c.~.}
When $\kappa$ is real and positive, this is precisely the same as the change in the low-energy effective action
of QCD when we add to the theory a quark mass proportional to $\kappa$ (the proportionality constant depends
on the chiral condensate). It leads to a pion mass equal to (this is sometimes called the
Gell-Mann-Oakes-Renner relation \GellMannRZ)
\eqn\pionmass{m_{\pi}^2 = {4 \kappa |\vev{OW}| \over f_{\pi}^2}~.}
Note that $\kappa$ and $f_{\pi}$ have dimensions of mass, while $OW$ has the
dimension of a mass cubed.

When $\kappa$ has an imaginary part (or is negative), the minimum of the pion
potential is no longer at $\pi(x)=0$ and the deformation \deform\ leads to a change in the phase
of the chiral condensate. We will assume that $\kappa$ is positive from here on (the other
cases are classically equivalent to this, since they are related by the axial $U(1)$ symmetry).

In addition to giving a mass to the pions, the perturbation \lamdeform\ changes the masses of the
massive mesons as well. To calculate their mass shifts, one needs to determine the shape of the $D8$-branes
in the presence of the perturbation. This shape is obtained by minimizing the deformed Lagrangian for
the eightbranes,
\eqn\defaction{L = T_{D8} \int dx^4 u^4 \sqrt{1 + \left({R_{D4}\over u}\right)^3 u'^2} +
2\, \kappa B \exp\left({1\over {2\pi \alpha'}} \int dx^4 u \right)~,}
where $B$ is the coefficient of the exponent in the computation of $\vev{OW}$, which is
necessary to give $L$ the appropriate dimension and to make the second term have the same scaling
${\cal O}(N_c)$ as the first term.\foot{This coefficient also depends on $u(x^4)$ through the
coupling of the worldsheet to the varying dilaton in our background. However, this dependence is
suppressed by a power of $\alpha'$ in the small curvature limit we are working in.} In the second
term in \defaction\ we used \areastraight.

The equation of motion corresponding to \defaction\ is
\eqn\defeom{{{T_{D8} R_{D4}^3} \over {\left[ 1 + \left({R_{D4}\over u}\right)^3 u'^2 \right]^{3/2}}}
\left[u u'' - {11\over 2} u'^2 - 4 \left({u\over R_{D4}} \right)^3 \right] =
{{\kappa B}\over {\pi \alpha'}} \exp\left({1\over {2\pi \alpha'}} \int dx^4 u \right)~.}
The right-hand side is independent of $x^4$, so the left-hand side is a constant. Denoting
this constant by $A \equiv \kappa |\vev{OW}| / \pi \alpha'$, \defeom\ is equivalent to the
first order differential equation
\eqn\defh{H = {{T_{D8} u^4}\over {\sqrt{1 + \left({R_{D4}\over u}\right)^3 u'^2}}} - A u = {\rm constant}~,}
associated with the symmetry $\{x^4\to x^4+{\rm constant}\}$ of \defaction. The constant value of $H$ may be
determined by requiring that $u$ goes to the UV cutoff $u=u_{\rm max}$ at
$x^4 = \pm L/2$. It is related to the minimal position $u_0$ of the $D8$-branes in the $u$ direction by
$H = T_{D8} u_0^4 - A u_0$.

Equation \defh\ enables us to compute the deformation in the position of the
$D8$-branes at leading order in $\kappa |\vev{OW}|$. This may then be used to determine the
shift of the
meson masses, by analyzing the quadratic fluctuations of the deformed action around this new
solution. It would be interesting to understand how to
go to higher orders in $\kappa$.

Note that, unlike the QCD mass deformation, the deformation \lamdeform\ in the strongly coupled $D4-D8-{\bar D}8$ theory
is irrelevant, and its effect grows in the UV region; this is clear from \defh. Thus, as for other
irrelevant deformations, the perturbation expansion in the deformation is only meaningful if we put in
a finite UV cutoff $u=u_{\rm max}$, and demand that the deformation is small at the cutoff scale.

It is easy to generalize the computations above to the Sakai-Sugimoto model in which
the $x^4$ direction is compactified. One interesting difference is that, in the special case of
$L = \pi R$, it seems natural to deform by the sum of the OWL operator \simpleowzero\ corresponding to
the contour connecting the $D8$ and $\bar D8$-branes in the positive $x^4$ direction, and
the one connecting them in the negative $x^4$ direction. In this case the shape of the
$D8$-branes is not modified by the deformation, since the two semi-classical strings
pull the $D8$-branes in opposite directions. Thus, in this special case, adding the
``quark mass deformation'' does not change the shape of the $D8$-branes, but it does give
a mass to the ``pions'' as discussed above. In all other cases, the shape of the $D8$-branes
is also modified; they are pulled to larger values of $u$ by the string.
In this model the distance between the minimal position of the $D8$-branes, $u_0$, and the minimal value of
the $u$ coordinate, $u_{\Lambda}$, may be interpreted as a constituent quark mass (at least in the context
of high-spin mesons \refs{\KruczenskiME\PeetersFQ-\BigazziJT}). We find generically (except for the
special case discussed above) that increasing the bare quark mass increases also the
constituent quark mass, as expected.

\newsec{$D$-branes in the background of $NS5$-branes}

In this section we study another example of a holographic description
of operators corresponding to long strings stretched between two $D$-branes.
This example is of interest for the study of $D$-brane dynamics near singularities
of the bulk geometry. It also has the advantage that the relevant classical string
background is under control, and can be analyzed exactly in $\alpha'$.

Consider the following brane configuration in type II string theory.
We start with $k$ $NS5$-branes stretched in $\IR^{5,1}$ labeled
by $(x^0,x^1,x^2,x^3,x^4,x^5)$, and located at the origin in the
transverse $\IR^4$. As is well known from the brane
construction of gauge theories (see \GiveonSR\ for a review),
$Dp$-branes which have one direction transverse to the
fivebranes can end on them. Thus, we add a $Dp$-brane stretched in the
directions $(x^0, x^1, x^2,\cdots, x^{p-1})$, and semi-infinite in
the $x^6$ direction (\ie\ it has $x^6\ge 0$ and ends on the fivebranes
at $x^6=0$).

\ifig\loc{The brane configuration : a $Dp$-brane and an $\bar Dp$-brane ending on $NS5$-branes.}
{\epsfxsize3.5in\epsfbox{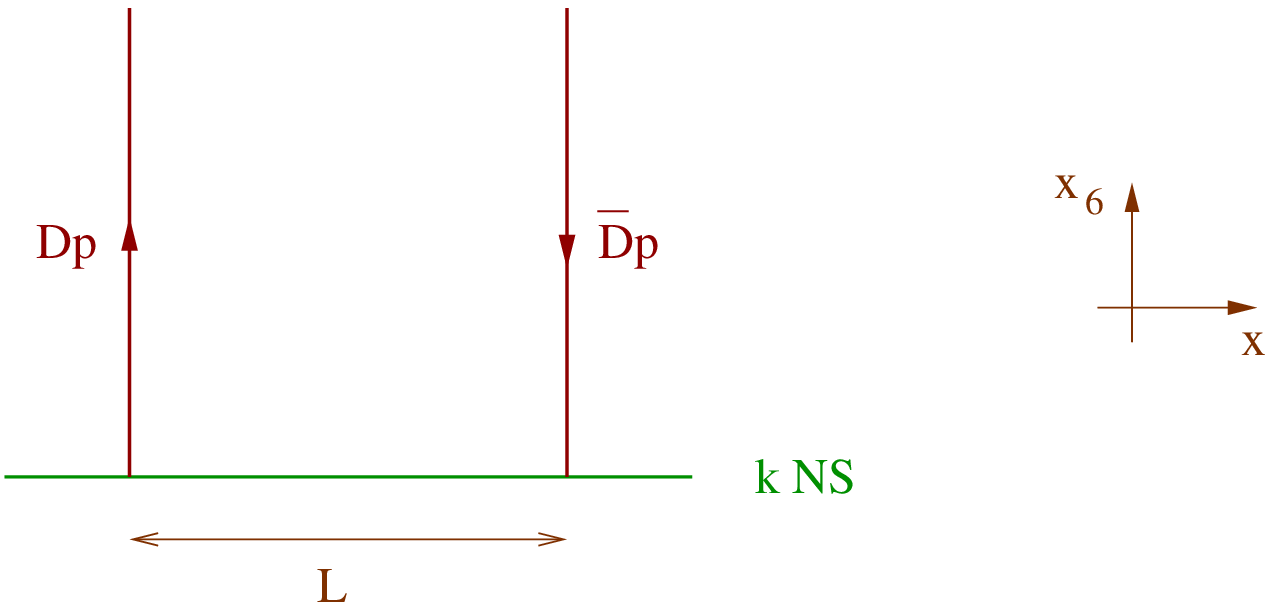}}

\noindent
The above $D$-brane is localized in the $\IR^{6-p}$ labeled by
$(x^p, x^{p+1}, \cdots, x^5)$. We can add a second $D$-brane, which
is parallel to the first one, but is displaced from it by a distance
$L$ in $\IR^{6-p}$, and has the opposite orientation, \ie\ it is a
$\bar Dp$-brane. We will label the direction along which the
$D$ and $\bar D$-brane are separated by $x$, with $x(D)=-{L\over2}$
and $x(\bar D)=+{L\over2}$. The brane configuration is depicted in \loc.

We will be primarily interested in the physics associated with the two
brane intersections in \loc. As reviewed in \GiveonSR, each
of the two intersections separately preserves 8 supercharges,\foot{The
system with both branes and anti-branes of course does not preserve any
supersymmetry.}  and carries no localized massless modes. One way to see
this is to compactify some of the directions along the fivebranes, and use
U-duality to turn each of the intersections in
the system in question to $k$ $D5$-branes stretched in
$(012345)$ intersecting a  $D3$-brane stretched in $(0126)$ along an $\IR^{2,1}$.
If the $D3$-brane is fully extended in $x^6$, $3-5$ strings give a massless
hypermultiplet in the fundamental representation of the low-energy $U(k)$ gauge symmetry on
the fivebranes, localized at the intersection. To reach the configuration of
interest to us, one needs to separate the two halves of the $D3$-brane (those with
positive and negative $x^6$) along the fivebranes, and to send the lower half to
infinity. This corresponds to giving an infinite mass to the hypermultiplet.

The endpoint of the $Dp$-brane on the fivebranes looks like a charged object
in the fivebrane theory. For example, for $p=1$, the $D1$-brane ending on the
$NS5$-branes gives rise to a static quark in the fundamental representation of
the low-energy $U(k)$ gauge theory of $k$ $NS5$-branes in type IIB string theory.
For $p=3$, the $D$-brane is extended in two of the directions along the fivebranes
$(12)$, and looks like a magnetic monopole in the remaining three.

While the system with just one intersection is uninteresting in
the infrared,\foot{In brane constructions of gauge theories, such systems do give interesting
infrared physics when embedded in richer brane configurations; this will not
play a role in our discussion below.} when both branes and anti-branes are
present, as in \loc, the situation is richer. Since we are
interested in the physics near the intersections, we can replace the fivebranes
by their near-horizon geometry, the CHS geometry \CallanAT:
\eqn\chs{ds^2=dx_\mu dx^\mu+d\phi^2+d\Omega^2~,}
where $\phi$ is related to the radial coordinate in the transverse $\IR^4$ as follows:
\eqn\defphi{r=g_s \sqrt{k\alpha'} \exp\left(\phi\over\sqrt{k\alpha'}\right)~,}
and $\Omega$ parameterizes the angular three-sphere in $\IR^4$, whose radius is given by
$\sqrt{k\alpha'}$. More precisely, the angular degrees of freedom are described by
a supersymmetric $SU(2)$ WZW model at level $k$.
$g_s$ is the asymptotic string coupling, far from the fivebranes. The geometry \chs\ is
obtained from the full fivebrane geometry by taking $g_s\to 0$ with $\phi$ held fixed;
in this limit it describes a ``little string theory'' (LST) (see \refs{\AharonyKS,\KutasovUF}
for reviews).
The dilaton behaves in this limit like
\eqn\dilbeh{\Phi=-{\phi\over\sqrt{k\alpha'}}~.}
A $Dp$-brane ending on the fivebranes corresponds in the geometry \chs\ -- \dilbeh\ to
a brane stretched in $(x^0,x^1,\cdots, x^{p-1},\phi)$, and localized on the three-sphere
and in $\IR^{6-p}$ \ElitzurPQ; the $\bar Dp$-brane is described similarly. As in the full
geometry, the $D$ and $\bar D$-branes are a distance $L$ apart in $\IR^{6-p}$. Note
that unlike the previous cases we discussed, here this distance does not grow as we
move out in the radial direction.

The $Dp$ and $\bar Dp$-branes attract each other via exchange of closed string modes,
but we will ignore this effect, and work just at leading order in the string coupling.
We will view the distance between the $Dp$ and
$\bar Dp$-branes at $\phi\to\infty$, $L$, as a fixed ($=$ non-normalizable) boundary
condition. Our focus here will be on the classical dynamics of normalizable open string
modes.

It turns out that for $L$ larger than a certain critical value,
\eqn\lcrit{L_{\rm crit}=\pi\sqrt{k\alpha'}~,}
all such modes are massive. As $L\to L_{\rm crit}$, a light mode appears.
For $L<L_{\rm crit}$ this mode becomes tachyonic and destabilizes the
brane configuration of \loc. A heuristic way of understanding this
instability is the following. The endpoints of the $Dp$ and $\bar Dp$-branes on
the $NS5$-branes attract each other via exchange of modes localized on the fivebranes
(LST modes). However, since the tension of the $Dp$-branes goes like the inverse
string coupling, while the attractive force due to exchange of a particular fivebrane
mode is of order one, this is a subleading effect in (the local) $g_s$.

A classical instability can only occur if the sum over the exchanges of all
modes of the LST diverges. Such a divergence can only be due to the contributions
of arbitrarily heavy LST states.  The contribution to the attractive force of a
given mode of mass $m$ decreases at large mass like $\exp(-mL)$, while the density
of LST states is well known to behave like $\rho(m)\sim\exp(2\pi\sqrt{\alpha' k}m)$.
Thus, superficially it seems that the sum over states diverges for $L<2L_{\rm crit}$ \lcrit.

This factor of two discrepancy is familiar from another, closely related, context --
closed string emission from accelerating branes in LST. It was argued in  \SahakyanCQ\
that it is natural to expect that the density of states that can be emitted by $D$-branes
in LST in fact goes like $\sqrt{\rho(m)}$. This would certainly be the case in ordinary
(critical) string theory, since a $D$-brane can only emit left-right symmetric closed
string states. Assuming that this is the case in LST as well, we conclude  that the
exchange of LST modes by the $D$-branes diverges precisely for $L<L_{\rm crit}$.

In the regime $k\gg 1$, $L_{\rm crit}$ is large in string units, and the above light mode
is best described as a translational mode of the $D$-brane configuration (which will be
described in detail below). For $k\sim 1$
or smaller\foot{Such values of $k$ in \dilbeh\ cannot arise in the near-horizon limit of flat
$NS5$-branes, but they can arise in other systems.}, a better description of this mode is as
a fundamental string stretched between the $D$ and $\bar D$-branes. We will consider the
geometric regime $k\gg 1$, but will return to this stretched string below.

To exhibit the geometric massless mode for $k\gg 1$, consider the projection of the
$D$-branes of \loc\ on the two dimensional space labeled by $(\phi,x)$. This
corresponds to a $D$-string described by a curve $x=x(\phi)$. The configuration of \loc\
corresponds to $x=\pm L/2$; the light mode corresponds to
deformations to a more general $x(\phi)$.  The DBI action for such a $D$-brane is given by
\eqn\actdp{S=-C\int dx \exp\left(\phi\over\sqrt{k\alpha'}\right)\sqrt{1+\phi'^2}~.}
Here $\phi=\phi(x)$, $\phi'=\partial_x\phi$, and $C$ is a known constant whose value will
not be needed below.

The fact that the Lagrangian \actdp\ does not depend explicitly on $x$ implies that one can
integrate the Euler-Lagrange equation once. After squaring the resulting equation one gets
\eqn\eomfirst{\exp\left(2\phi\over\sqrt{k\alpha'}\right)=1+\phi'^2~,}
where we fixed a constant that appears in the integration to a particular value by shifting $\phi$.
The solution of \eomfirst\ is
\eqn\hairp{\exp\left(-{\phi\over\sqrt{k\alpha'}}\right)=\cos\left(x\over\sqrt{k\alpha'}\right)~.}
\ifig\hairpin{The hairpin $D$-brane \hairp.}
{\epsfxsize4.5in\epsfbox{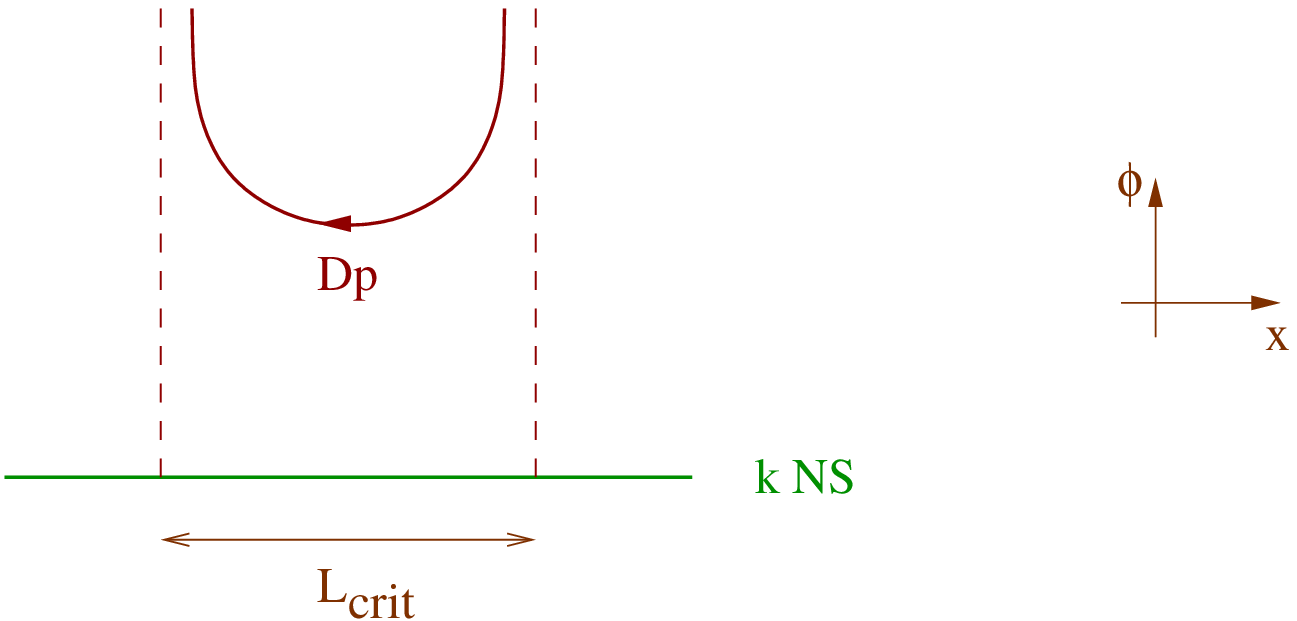}}
\noindent
It describes a U-shaped connected brane, the hairpin brane of \LukyanovNJ\ (or, more precisely,
its generalization to the fermionic string discussed in \refs{\KutasovDJ,\NakayamaYX} and other
papers). As $\phi\to\infty$, it approaches a brane and anti-brane a distance $L_{\rm crit}$ \lcrit\
apart. As $\phi$ decreases, the two $D$-branes bend towards each other; they smoothly connect at
$\phi=0$ (see \hairpin).

As mentioned above, the position of the bottom of the brane depicted in \hairpin\
is a free parameter of the solution, as is clear from the form of the action \actdp.
Moreover, the energy of the brane is independent of this parameter.
Thus, when the distance between the $D$ and $\bar D$-branes at infinity is
equal to the critical one \lcrit, the mode corresponding to fluctuations of the bottom
of the hairpin brane is massless and has a flat potential (at leading order in the
string coupling).

When $L>L_{\rm crit}$, this mode is massive, and the hairpin tends to collapse back to the
brane-antibrane configuration of \loc. For $L<L_{\rm crit}$ it is tachyonic and
the bottom of the U-shape tends to run to large $\phi$. The resulting time-dependent
solutions can be described using techniques similar to those of  \ItzhakiZR, who studied
a closed string analog of this problem.

The original brane configuration of \loc\ has a $U(1)\times U(1)$ symmetry associated
with the two $Dp$-branes. This is a local symmetry on the $D$-branes, but from the
point of view of the LST it is a global one. This symmetry is broken to
the diagonal $U(1)$ when the branes connect. It is interesting to ask whether there is an
operator that is charged under the broken $U(1)$ and has a non-zero expectation value
in the configuration of \hairpin. Such an operator could serve as an order parameter for
the symmetry breaking described above, as in our discussion of the $D4-D8-{\bar D}8$
system in section 3.

A natural candidate for such an operator is a string stretched between the $Dp$ and
$\bar Dp$-branes in \loc. The lowest lying state of such a string is the open string
``tachyon'' stretched between the two branes. From studies of the hairpin brane, which turns
out to be described by an exactly solvable boundary conformal theory, it is known that
such an operator is indeed turned on in the vacuum. In the bosonic string this was discussed
in \refs{\LukyanovNJ,\LukyanovBF}, while in the fermionic case of interest to us here in \KutasovRR.

Asymptotically, at large $\phi$, the worldsheet Lagrangian contains a term corresponding to
a boundary ${\cal N}=2$ superpotential, which behaves like
\eqn\boundsup{\delta S_{\rm ws}=\mu\int dtd\theta
\exp\left[-\half\sqrt{k\over\alpha'}(\phi+i\tilde x)\right]+c.c.~.}
Here $\tilde x=x_L-x_R$ is the T-dual of $x$, and the coupling $\mu$ is determined by $\phi_{IR}$,
the location of the bottom of the hairpin brane. The dependence can be determined by a scaling
argument of the kind familiar from Liouville theory. For the hairpin shape \hairp, one has
$\phi_{IR}=0$, which corresponds to some particular $\mu=\mu^{(*)}$. If we replace $\phi\to\phi-\phi_{IR}$,
such that the bottom of the hairpin is at $\phi=\phi_{IR}$, we see from \boundsup\ that
\eqn\mmuu{\mu=\mu^{(*)}\exp\left(\half\sqrt{k\over\alpha'}\phi_{IR}\right)~.}
When $\phi_{IR}\to -\infty$, the bottom of the hairpin brane descends into the strong
coupling region and one smoothly approaches the parallel brane-antibrane configuration
of \loc. $\mu$ \mmuu\ also goes to zero in this limit.\foot{When $\phi_{IR}$ becomes too small,
we cannot trust the shape of the bottom of the hairpin due to strong quantum effects, but there
is no reason to expect non-smooth behavior there.}

As mentioned above, the large symmetry of the problem (${\cal N}=2$ worldsheet superconformal symmetry) allows
one to solve the boundary conformal field theory corresponding to the hairpin brane exactly, and in particular
one can deduce the presence of the boundary ${\cal N}=2$ superpotential \boundsup. Thus, it is interesting to
study this case in detail, in the hope of developing techniques which could be useful also in more general
circumstances where the worldsheet theory is not solvable, such as backgrounds with Ramond-Ramond fields turned on.

In particular, we would like to understand the origin of \boundsup\ at large $k$, where both
the closed string background \chs\ -- \dilbeh, and the shape of the $D$-brane \hairp, are
slowly varying, and we can expect semi-classical techniques to be valid. To do that it
is useful to note that the boundary superpotential \boundsup\ is a normalizable operator
at large $\phi$. As is familiar from holography in general, $\mu$ is proportional to the
expectation value of the non-normalizable operator that creates a string stretched between
the $D$ and $\bar D$-branes at the boundary. This operator, which is analogous to the OWL
operators described in the previous sections, behaves at large $\phi$ like
\eqn\nonnormt{T\simeq
\exp\left[\left({1\over2}\sqrt{k\over\alpha'}-{1\over\sqrt{k\alpha'}}\right)\phi-
{i\over2}\sqrt{k\over\alpha'}\tilde x\right]~.}
Thus, we need to calculate the expectation value of \nonnormt\ in the hairpin state. A
scaling argument similar to that described above implies that if this expectation value
is non-zero, it is indeed proportional to $\mu$ \mmuu.

To calculate this expectation value it is useful to note that the tachyon background
\boundsup\ is a non-perturbative effect in the worldsheet theory, whose loop expansion
parameter is $1/\sqrt k$ (the curvature of the $D$-brane).
Thus, it is natural to expect that it is due to a worldsheet instanton
effect, involving an open string ending on the boundary; this also follows from our
general discussion in the previous sections of the holographic dual of long open
strings. The instanton in question is a map from the worldsheet disk $|z|\le 1$ to
the part of the two dimensional $(x,\phi)$ plane bounded by the hairpin,
\eqn\boundhair{\exp\left(-{\phi\over\sqrt{k\alpha'}}\right)\le\cos \left(x\over\sqrt{k\alpha'}\right)~.}
Near the boundary $\phi\to\infty$ this worldsheet looks like a string stretched between
the $D$-branes, which implies that this configuration contributes to the one point function
of the stretched string operator \nonnormt.

The instanton configuration can be constructed as follows. Start with the worldsheet action
\eqn\caca{S={1\over\pi\alpha'}\int d^2z (\partial_z\phi\partial_{\bar z}\phi
+\partial_zx\partial_{\bar z}x)~.}
It is convenient to parameterize the $(x,\phi)$-plane by the coordinate
\eqn\ddd{U=\exp\left(\phi-\phi_{IR}+ix\over\sqrt{k\alpha'}\right)~,}
in terms of which the hairpin shape \hairp\ takes the simple form
\eqn\eee{U+U^*=2~,}
or, equivalently, ${\rm Re}(U-1)=0$.

The worldsheet action \caca\ now takes the form
\eqn\fff{S={k\over\pi}\int d^2z {1\over |U|^2}
(\partial_zU\partial_{\bar z} U^*+\partial_{\bar z} U\partial_z U^*)~.}
The disk instanton we are looking for is a holomorphic map from the disk to the
half-plane bounded by \eee, and is easy to write down:
\eqn\ggg{U-1={1+z\over 1-z}~.}
Its action is proportional to the area $A$ of the Euclidean string worldsheet \ggg:
\eqn\hhh{S_{\rm inst}={A\over2\pi\alpha'}~.}
This area is infinite, since as $\phi\to\infty$ the hairpin looks like two
D-strings a distance $L_{\rm crit}$ \lcrit\ apart, so there is a divergence
from that region. This divergence can be regulated by introducing an
upper bound on $\phi$, $\phi_{UV}$, which can be thought of as a UV cutoff.

In any case, we are only interested in the dependence of the area on the
position of the bottom of the hairpin, $\phi_{IR}$, discussed around
\mmuu. We can isolate this dependence by differentiating the area with respect to
$\phi_{IR}$. A short calculation leads to
\eqn\diffarea{{\partial A\over\partial\phi_{IR}}=-L_{\rm crit}}
in the limit $\phi_{UV}\to\infty$. Therefore, after rescaling the operator
\nonnormt\ by a factor which depends on the UV cutoff, we conclude that
\eqn\vevtt{\langle T\rangle\sim \exp\left(-S_{\rm inst}\right)\sim
\exp\left(L_{\rm crit}\phi_{IR}\over 2\pi\alpha'\right)\sim \mu~,}
where we used \lcrit, \mmuu, \hhh, \diffarea. We see that indeed the instanton contribution
scales in the right way with $\phi_{IR}$ to give a non-zero one-point function
to the long open string operator \nonnormt. Note that we have only computed the
leading exponential contribution to the one point function. The pre-exponential
factor involves contributions from the dilaton coupling in the worldsheet action,
and the determinant of small fluctuations around the instanton \ggg. These are
subleading in the large $k$ limit, and are expected to give rise to a
constant contribution to \vevtt\ (independent of $\phi_{IR}$).

So far we discussed the spontaneous breaking of the $U(1)\times U(1)$ symmetry
of the brane configuration of \loc\ to the diagonal $U(1)$, by the brane configuration
of \hairpin. We have seen that the order parameter for this breaking can be taken
to be the stretched string operator \nonnormt, and it indeed has a non-zero
expectation value in the hairpin state \vevtt.
It is natural to ask what happens if we deform the system by adding to the worldsheet
Lagrangian the non-normalizable operator $T$ \nonnormt,
\eqn\defws{\delta S_{ws}={\kappa\over2}\int dtd\theta T(\phi,x)+c.c.~.}
This deformation breaks the $U(1)\times U(1)$ symmetry explicitly. It also breaks
the ${\cal N}=2$ superconformal symmetry of the hairpin brane; therefore we do not expect the
resulting theory to be exactly solvable. However, one can still ask how the shape of the
$D$-brane and its low-lying spectrum change in the presence of this deformation.

To first order in $\kappa$ and in the semi-classical regime  $k\gg 1$ one can answer this
question by adding to the DBI action \actdp\ the exponential of the Nambu-Goto action
for the instanton string discussed above,
\eqn\modact{S= -C\int dx \exp\left(\phi\over\sqrt{k\alpha'}\right)\sqrt{1+\phi'^2}
-\kappa Be^{-S_{\rm NG}}~.}
Here, as before (see \hhh), $S_{NG}={A\over2\pi\alpha'}$, where $A$ is the area
of a minimal worldsheet enclosed by the deformed hairpin, and $B$ is the pre-exponential
factor in the expectation value of $T$ above. It depends on the shape of the deformed
hairpin, but for the purpose of the calculation below, to leading order in the  $1/k$ expansion
we can neglect this dependence.

To calculate the shape of the deformed hairpin to first order in $\kappa$ we need to
solve the equation of motion of $\phi(x)$ with the deformed action \modact. For this
we need the dependence of $S_{\rm NG}$ on the shape $\phi(x)$. It is easy to see that
it is given by
\eqn\ssmm{S_{\rm NG}=-{1\over2\pi\alpha'}\int dx\phi(x)+\cdots~,}
where the ellipsis stand for terms that depend on the UV cutoff $\phi_{UV}$, but not
on the shape $\phi(x)$. Varying \modact\ with respect to $\phi(x)$ and integrating once,
we find the first order equation
\eqn\firstmod{C{e^{\phi\over\sqrt{k\alpha'}}\over \sqrt{1+\phi'^2}}+
{\kappa\langle T\rangle\over2\pi\alpha'}\phi=D~,}
where $D$ is a function of $\phi_{IR}$ (or, equivalently, of the separation between
the brane and anti-brane at some UV cutoff $\phi_{UV}$).
This equation generalizes \eomfirst\ to non-zero $\kappa$, and it can be solved by expanding
$\phi$ as $\phi=\phi_0+\kappa\phi_1+\cdots$, and keeping only first order terms in $\kappa$.
For example, at large $\phi$, the leading deformation of the hairpin from its original form
is given by
\eqn\deformh{C\partial_\phi x=\left( D-{\kappa\langle T\rangle\over2\pi\alpha'}\phi\right)
e^{-{\phi\over\sqrt{k\alpha'}}}~.}
Of course, when $\phi$ becomes too large, one has to go beyond the linear approximation in $\kappa$
described above.

It is interesting to compare the deformed shape of the hairpin \deformh\ which we
found above, to the deformed shape implied by the effective action on the $Dp$-brane
coupled to the ``tachyon'' field $T$. In curved space and for curved $D$-branes (as in the
discussion of the previous sections) it is not known how to write down such an effective
action, but for flat $D$-branes in flat space we know how to write it down, and this is
the situation in the asymptotic region of the hairpin. In this region we know that, if
we denote the distance between the brane and the anti-brane by $L_{\rm crit}-2x(\phi)$ (where
$x(\phi)$ is small in the UV), the mass of the open string ground state is given by
\eqn\openstringmass{m^2(x(\phi)) = -{1\over {2\alpha'}} + \left({{L_{\rm crit}-2x} \over {2\pi \alpha'}}
\right)^2 \simeq m_0^2 -  \sqrt{k\over \alpha'^3} {x\over\pi}~.}

\noindent
The effective action of the ``tachyon'' stretched between the $D$ and $\bar D$-branes in \hairpin\ is
given to quadratic order by
\eqn\nactdp{S = - C \int d\phi \exp\left({\phi \over \sqrt{k \alpha'}}\right) \left[(\del_{\phi} x)^2 +
(\del_{\phi} T)^2 + m^2(x) T^2\right]~.}
We are looking for a configuration where the normalizable mode of the ``tachyon''
\boundsup\ is turned on with a coefficient $\vev{T}$, and the non-normalizable
mode \nonnormt\ is turned on with a coefficient $\kappa$, such that at leading
order in $\kappa$ the tachyon field behaves asymptotically as
\eqn\asymt{T^2 \simeq \beta \kappa \vev{T} \exp(-{\phi \over \sqrt{k \alpha'}})~,}
where $\beta$ is a constant coming from carefully normalizing the normalizable and
non-normalizable modes of the ``tachyon''.

The equation of motion of $x(\phi)$ with this ``tachyon'' source, at leading order in $\kappa$ (and in
the UV region where $x$ is small), then takes the form
\eqn\xeom{\del_{\phi} \left[\exp\left({\phi \over \sqrt{k \alpha'}}\right) \del_{\phi} x\right] = -{\beta
\kappa \vev{T} \over {2\pi \alpha'}} \sqrt{k\over \alpha'}~.}
For $1/\beta = C \sqrt{k /\alpha'}$, this precisely agrees with \deformh\ above.

\newsec{Additional issues}

\subsec{A supersymmetric example}

The main examples we focused on so far were non-supersymmetric, but one can also
construct interesting examples of OWL operators \openwilson\ in
supersymmetric theories, including examples which preserve some of the supersymmetry.
We will describe here just one example, leaving a further investigation to future
work.

Consider the $d=4$ ${\cal N}=4$ $SU(N_c)$ SYM theory coupled to $N_f$ three dimensional massless hypermultiplets living on the surface $x_3=0$. In the 't Hooft large $N_c$
limit with 't Hooft
coupling $\lambda_4$ and with fixed $N_f$, this is described by type IIB string theory
on $AdS_5\times S^5$, with $N_f$ $D5$-branes filling an $AdS_4\times S^2$ subspace
\refs{\KarchGX,\DeWolfePQ}; if we
use the Poincar\'e coordinates of $AdS_5$ (with a boundary at $z\to 0$),
\eqn\poincare{ds^2 = \sqrt{\lambda_4} \alpha' {{dx_{\mu}^2 + dz^2} \over {z^2}}~,}
then the D5-branes are simply located at $x_3=0$ (and wrap some maximal $S^2$ inside the
$S^5$). This theory breaks half of the supersymmetry of the ${\cal N}=4$ SYM theory; it
preserves a $d=3$ ${\cal N}=4$ superconformal symmetry.

Now, consider an OWL starting at a hypermultiplet at
$x_0=x_1=x_2=x_3=0$ and stretching to infinity in the $x_3$ direction. Such an operator
is analogous to the ``straight Wilson line'' in the ${\cal N}=4$ SYM theory; it is well-defined
if we put appropriate boundary conditions at infinity. In the holographic dual
description, the computation of the one-point function of this operator is dominated by
a string sitting at $x_0=x_1=x_2=0$ and filling the
$z$ axis and the positive $x_3$ axis in \poincare\ (we assume that the OWL
couples to a scalar such that the string lives at a point in the $S^2$ filled by the
$D5$-branes). This operator breaks half of the
supersymmetry (leaving 8 unbroken supercharges, including both regular supercharges
and superconformal charges), and the holographic computation of its VEV gives one, since the regularized area of the surface vanishes (just like for
the ``straight Wilson line'').

This case is not very interesting, but suppose that we now perform a conformal transformation
involving an inversion around a point $x_0=x_1=x_3=0$, $x_2=a$. This transformation
leaves the field theory described above invariant.
However, the contour in the OWL now maps to a semi-circle
\eqn\semicircle{(x_2-a+{1\over {2a}})^2 + x_3^2 = {1\over {4a^2}}~,\qquad x_3 \geq 0~.}
This is a standard OWL connecting two hypermultiplets
of the form \openwilson, with a semi-circular contour \semicircle\ between the two
points $(x_2=a,x_3=0)$ and $(x_2=a-1/a, x_3=0)$.
Our derivation of this configuration by a conformal
transformation ensures that this OWL still
preserves 8 supercharges, though these are now combinations of standard
supersymmetries and superconformal symmetries.

The holographic computation of the one-point function of this OWL is straightforward; the dominant solution is
just half of the solution for the circular Wilson line \refs{\BerensteinIJ,\DrukkerZQ}, with a string worldsheet at
\eqn\semicircle{(x_2-a+{1\over {2a}})^2 + x_3^2 + z^2 = {1\over {4a^2}}~,\qquad x_3 \geq 0~.}
Its area is thus half of that corresponding to the circular Wilson line, which is
$\sqrt{\lambda_4}$, so
the VEV of the OWL (at leading order in the $\alpha'$ expansion) is
equal to $\exp(\sqrt{\lambda_4}/2)$.

In the case of the closed circular Wilson line
case it has been conjectured \refs{\EricksonAF,\DrukkerRR} and recently proven \PestunRZ\ that the result is given
by a zero-dimensional matrix model, since the conformal transformation can only change
the result because a point is brought in from infinity.
Similar arguments imply that the semi-circular open Wilson line
$\vev{OW}$ described
above should also be computable by a zero-dimensional model of matrices and vectors; it
would be interesting to verify this.

\subsec{Divergences in open Wilson line computations}

Closed supersymmetric Wilson loops are known to have divergences at cusps, which can be
computed both perturbatively and at strong coupling (with a qualitatively similar behavior found in both
limits \DrukkerZQ). Similarly, in the case that the fields in the fundamental representation are localized
on some subspace, the correlation functions of the
open Wilson line observables \openwilson\ have a divergence whenever
the contour $\tilde C$ ends on that subspace at an angle
which is not a straight angle. In this section we describe this divergence both at weak coupling
(using perturbation theory) and at strong coupling (using the mapping to string worldsheets).

Let us consider a $D$-dimensional large $N$ gauge theory, in which some fields in the fundamental
representation are localized on a $d$-dimensional subspace; without loss of generality we can take this subspace to be
\eqn\subspace{x^{d+1}=x^{d+2}=\cdots=x^D=0~.}
When we consider an open Wilson line operator of the
form \openwilson, starting at a fundamental field located at $x=0$, there is now an angle
associated with this operator, which is the angle $\theta$ between the direction of the Wilson
line (near $x=0$) and the subspace that the fundamental fields live on. For instance, again without loss of
generality, we can assume that near $x=0$ the Wilson line (parameterized by $t$) looks like
\eqn\contour{x^{d+1} = t \sin(\theta),\qquad x^d = t \cos(\theta),\qquad t \geq 0~,}
implying that the Wilson line couples
to the gauge field components $\sin(\theta) A_{d+1} + \cos(\theta) A_d$. On the other hand,
the fields in the fundamental representation couple just to $A_d$ and they do not couple
to $A_{d+1}$. The one-loop diagram involving the exchange of a gauge field between the Wilson
line and the propagator of the field in the fundamental representation
then has a divergence as $t\to 0$, proportional (near
$\theta = \pi/2$) to $\cos^2(\theta)$, going as $\int dt / t^{D-3}$. For $D=4$ we have a
logarithmic divergence (as for a cusp in a closed Wilson line), and for $D=5$ a linear
divergence. The only case in which there is no divergence is when the Wilson line intersects
the surface \subspace\ at a straight angle $\theta=\pi/2$.

When the fundamental representation fields couple also to scalar fields (note that this is not the case
in the $D4-D8-{\bar D8}$ system), then, for a specific
choice of the scalar field couplings of the open Wilson line, it may be possible to cancel
this divergence. However, generally this divergence is present even for locally supersymmetric OWL operators (as is the case for the cusp divergence).

On the strong coupling side, for the purposes of computing the divergence in the open Wilson
line correlators we can concentrate just on the
region near the boundary, where the flavor $D$-brane just sits at $x^{d+1}=\cdots = x^D=0$ and
stretches in the radial direction. We need to find a minimal worldsheet ending on the contour
\contour\ at the boundary and transverse to the $D$-brane.
It is easy to convince oneself that such a worldsheet
is the same as half of the closed string worldsheet ending on the contour
\eqn\contourtwo{x^{d+1} = t \sin(\theta),\qquad x^d = |t| \cos(\theta)~,}
that we obtain by joining to \contour\
its reflection around the subspace that the fundamental fields live on (a similar trick
was recently used in \McGreevyKT). This contour has a
cusp at $t=0$ with an angle of $2\theta$, so it leads to a divergence which is similar to the cusp
divergence occurring in closed Wilson loops (whenever $\theta \neq \pi/2$). For the case of
$D=4$ this is a logarithmic divergence, just as in the previous paragraph, but its precise
dependence on the angle is different from the one found at weak coupling (this is also true
for the closed Wilson loop cusp divergence) \DrukkerZQ.

Note that in this computation we assumed that the end of the open Wilson line is at the same
position as the $D$-brane in the compact directions (otherwise there is no semi-classical worldsheet
contributing to the computation of correlation functions of $OW$).
If the $D$-brane is partially localized in the compact
directions (so that the fundamental fields couple to some of the scalar fields of the gauge theory)
then this implies that near the end of the open Wilson line, the Wilson line couples to different scalar
fields than the ones which the fundamental fields couple to. Thus, for such Wilson lines there
is no contribution from the scalar fields at leading order in perturbation theory,
and their one-loop computation diverges as described above.

In any case, we showed that both at weak coupling and at strong coupling, when the fundamental representation fields
are localized on a subspace, one has to choose
the Wilson line operators \openwilson\ such that the direction of the open Wilson line is transverse to that
subspace at its beginning and end, in order to avoid cusp-like divergences in the computation.

\vskip 1cm
\centerline{\bf Acknowledgements}

We would like to thank O. Bergman, N. Drukker, S. Hartnoll, Z. Komargodski, O. Lunin,
D. Reichmann, A. Schwimmer, J. Sonnenschein, and S. Yankielowicz for useful discussions.
The work of OA was supported in part by the
Israel-U.S. Binational Science Foundation, by a center of excellence supported by the Israel Science Foundation
(grant number 1468/06), by a grant (DIP H52) of the German Israel Project Cooperation, by the European network MRTN-CT-2004-512194, and by Minerva.
DK is supported in part by DOE grant DE-FG02-90ER40560, by the
National Science Foundation under Grant 0529954, and by the
Israel-U.S. Binational Science Foundation. DK thanks the Weizmann
Institute for hospitality during part of this work.

\listrefs

\end